\documentclass[pra,twocolumn,amsmath,amssymb,superscriptaddress,longbibliography,eqsecnum]{revtex4-1}
\usepackage{graphicx,amsmath,relsize,epstopdf,upgreek,color,mathtools,bm,mathptmx}
\usepackage[colorlinks=true,linkcolor=blue,citecolor=blue,urlcolor =blue]{hyperref}

\newcommand{\ket}[1]{\left|{#1}\right>}
\newcommand{\bra}[1]{\left<{#1}\right|}
\newcommand{\rvec}[1]{\pmb{#1}}
\newcommand{\dyadic}[1]{\pmb{#1}}

\newcommand{\D}{\mathrm{d}}
\newcommand{\I}{\mathrm{i}}
\newcommand{\TP}[1]{{#1}^\mathrm{\,\textsc{t}}}
\newcommand{\E}[1]{\mathrm{e}^{\mbox{\footnotesize$#1$}}}
\newcommand{\Tr}[1]{\mathrm{Tr}\!\left\{#1\right\}}
\newcommand{\csch}{\mathrm{cosech}\,}

\newcommand{\DET}[1]{\det\!\left\{#1\right\}}
\newcommand{\RE}[1]{\mathrm{Re}\!\left\{#1\right\}}

\newcommand{\LAG}[2]{\mathrm{L}_{\,#1}\!\left(#2\right)}
\newcommand{\VAR}[2]{\mathrm{Var}_{#1}\!\left[#2\right]}
\newcommand{\MEAN}[2]{\mathbb{E}_{#1}\!\left[#2\right]}

\newcommand{\FHOMoneN}{\widetilde{\dyadic{F}}_{1,\textsc{hom}}}
\newcommand{\FHOMtwoN}{\widetilde{\dyadic{F}}_{2,\textsc{hom}}}

\newcommand{\VEC}[1]{\mathrm{vec}\!\left(#1\right)}

\newcommand*\pFqskip{8mu}
\catcode`,\active
\newcommand*\pFq{\begingroup
	\catcode`\,\active
	\def ,{\mskip\pFqskip\relax}%
	\dopFq
}
\catcode`\,12
\def\dopFq#1#2#3#4#5{%
	{}_{#1}\mathrm{F}_{#2}\!\left({#3};{#4};#5\right)%
	\endgroup
}

\begin{document}

\title{When heterodyning beats homodyning: an assessment with quadrature moments}

\author{Y.~S.~Teo}
\affiliation{BK21 Frontier Physics Research Division, 
	Seoul National University, 08826 Seoul, South Korea}

\author{C.~R.~M\"{u}ller}
\affiliation{Max-Planck-Institut f\"ur  die Physik des Lichts,
	Staudtstra\ss e 2, 91058 Erlangen, Germany}

\author{H.~Jeong}
\affiliation{Center for Macroscopic Quantum Control,
	Seoul National University, 08826 Seoul, South Korea}

\author{Z.~Hradil}
\affiliation{Department of Optics, Palack\'{y}  University,
	17. listopadu 12, 77146 Olomouc, Czech Republic}

\author{J.~\v{R}eh\'{a}\v{c}ek}
\affiliation{Department of Optics, Palack\'{y}  University,
	17. listopadu 12, 77146 Olomouc, Czech Republic}

\author{L.~L.~S\'{a}nchez-Soto}
\affiliation{Departamento de \'Optica, Facultad de F\'{\i}sica,
	Universidad Complutense, 28040 Madrid, Spain}
\affiliation{Max-Planck-Institut f\"ur  die Physik des Lichts,
	Staudtstra\ss e 2, 91058 Erlangen, Germany}

\begin{abstract}
  We examine the moment-reconstruction performance of both the homodyne and heterodyne (double-homodyne) measurement schemes for arbitrary quantum states and introduce moment estimators that optimize the respective schemes for any given data. In the large-data limit, these estimators are as efficient as the maximum-likelihood estimators. We then illustrate the superiority of the heterodyne measurement for the reconstruction of the first and second moments by analyzing Gaussian states and many other significant non-classical states.
\end{abstract}


\maketitle

\section{Introduction}

The next-generation quantum technologies introduce novel and innovative routes to the understanding and implementation of measurements, communication, and computation. In this respect, the manipulation of a quantum light source using continuous-variable~(CV) measurements offer many advantages~\cite{Braunstein:2005aa,Ferraro:2005ns,CV2007:aa,Andersen:2010ng,Adesso:2014pm}. There exist two standard $CV$ measurement schemes. The more commonly employed homodyne detection~\cite{Yuen:1983ba,Abbas:1983ak, Schumaker:1984qm}, which performs an approximate measurement of rotated photonic quadratures~\cite{Banaszek:1997ot}, probes the marginal distribution of the Wigner function of the unknown quantum state~\cite{Vogel:1989zr}. The other less widely adopted double-homodyne detection, or the heterodyne detection, involves the joint measurement of complementary observables~\cite{Arthurs:1965al,Yuen:1982hh,Arthurs:1988aa,Martens:1990al,Martens:1991aa,Raymer:1994aj,Trifonov:2001up,Werner:2004as}that directly samples the phase space according to the Husimi function~\cite{Stenholm:1992ps} and is connected to the conventional heterodyne scheme~\cite{Javan:1962aa,Read:1965aa,Carleton:1968aa,Gerhardt:1972aa,Yuen:1980ys,Shapiro:1984aa,Shapiro:1985aa,Walker:1986qn,Collett:1987aa,Lai:1989ah}.

These measurement schemes, which experimentally probe quasi-probability distributions, can also be equivalently understood as practical means to directly characterize the source in terms of the ordered moments of the quadrature operators in phase space. Gaussian states~\cite{Ferraro:2005ns} for example, which are important in analyzing CV quantum information processing~\cite{Lorenz:2004aa, Lance:2005aa,Scarani:2009cq,Weedbrook:2012ag}, are conveniently described by this representation since all their operator moments are functions of only the first and second moments. Therefore, estimating the first and second moments are enough to fully reconstruct the Gaussian state or verify if the reconstructed state is accurately Gaussian \cite{Rehacek:2009ys}. Higher moments come into play for general quantum states. On its own right, the topic of operator moments of quantum states draws interest in the context of generalized uncertainty relations~\cite{Angulo:1993aa,Angulo:1994ps}, non-classicality detection~\cite{Simon:9709030,Arvind:1998aa}, entanglement detection~\cite{Namiki:2012gs,Ivan:2012da}, and cryptography~\cite{Leverrier:2012qs,Thearle:2016aa}.

In Refs.~\cite{Rehacek:2015qp} and \cite{Muller:2016da}, we theoretically and experimentally compared the two measurement schemes, using a polarization-squeezing setup~\cite{Heersink:2005ul,Grangier:1987fk,Josse:2004ys,Marquardt:2007bh,Muller:2012ys,Peuntinger:2014aa} for the latter. We analyzed the physical implications of having the unavoidable Arthurs-Kelly type noise that is inherent in the joint measurement heterodyne scheme on moment reconstruction. We found that despite this additional noise, for a single-mode central-Gaussian source the heterodyne scheme still results in second-moment estimators that are more accurate than the homodyne scheme for a wide range of the squeezing strength and temperature parameter.

In this article, we extend the theory of these two CV measurement schemes to general quantum states and show that the tomographic advantage in using the heterodyne scheme carries over to other interesting and important non-Gaussian states. This message is conveyed in five main sections. Section~\ref{sec:cov_mom} gives an overview of the fundamental elements in first- and second-moment tomography, as well as the concept of reconstruction accuracy. These elements are then used to discuss the general theory of moment reconstruction for the homodyne and heterodyne schemes in Sec.~\ref{sec:gen_theory}. In that section, we shall also introduce optimal moment estimators that asymptotically approach the respective Cram{\'e}r-Rao bounds, which are derived in Appendix~\ref{app:optimality}. In Sec.~\ref{sec:first-mom}, we shall study the CV schemes in first-moment estimation where it shall be shown that heterodyne detection will always outperform homodyne detection unless the state is of minimum uncertainty, in which case the two schemes give equal reconstruction accuracy per sampling event. This result shall be discussed with some interesting classes of non-Gaussian states. Next, we study the results for second-moment estimation Sec.~\ref{sec:sec-mom} with the same classes of non-Gaussian states and illustrate once again the tomographic advantages of using the heterodyne scheme in moment tomography. Finally, Sec.~\ref{sec:conc} concludes the presented results in a summary.

\section{The covariance matrix and moment-reconstruction accuracy}
\label{sec:cov_mom}

In dealing with single-mode bosonic systems such as photons, for the pair of position $X$ and momentum $P$ quadrature operators obeying $[X,P]=\I$ (with the quantum unit $\hbar\equiv1$) that form the column $\rvec{R}=\TP{(X\,\,\,P)}$, the covariance matrix can be written as 
\begin{equation}
\dyadic{G}=\RE{\langle\rvec{R}\rvec{R}^\textsc{t}\rangle}-\langle\rvec{R}\rangle\TP{\langle\rvec{R}\rangle}=\dyadic{G}_2-\dyadic{G}_1\,,
\end{equation}
where we have introduced the first- ($\dyadic{G}_1=\langle\rvec{R}\rangle\TP{\langle\rvec{R}\rangle}$) and second-moment ($\dyadic{G}_2=\RE{\langle\rvec{R}\rvec{R}^\textsc{t}\rangle}$) matrices. The two independent parameters $\left\{\langle X\rangle,\langle P\rangle\right\}$ in $\dyadic{G}_1$ and three independent parameters $\left\{\langle X^2\rangle,\frac{1}{2}\langle\{X,P\}\rangle,\langle P^2\rangle\right\}$ in $\dyadic{G}_2$ constitute the complete set of five parameters that characterize $\dyadic{G}$. The well-known class of Gaussian states possesses a Gaussian Wigner function or any other kind of well-behaved quasi-probability distribution. As a consequence, any Gaussian state is fully described by only $\dyadic{G}_1$ and $\dyadic{G}_2$.

The covariance matrix for any quantum state obeys the inequality $\dyadic{G}\geq\dyadic{\sigma}_y/2$ in terms of the Pauli matrix $\dyadic{\sigma}_y$, which is a recast of the Heisenberg-Robertson-Schr{\"o}dinger (HRS) uncertainty relation for position and momentum operators. This gives the equivalent stricter inequality $\DET{\dyadic{G}}\geq1/4$ in addition to the standard positivity constraint for $\dyadic{G}$. The reconstruction of the full covariance matrix $\dyadic{G}$ involves the quantum tomography of all the five independent parameters that define the first and second operator moments of the state. Here, the figure of merit the reconstruction accuracy is the mean squared-error~(MSE) $\mathcal{D}=\MEAN{}{\Tr{\left(\widehat{\dyadic{G}}-\dyadic{G}\right)^2}}$ between $\dyadic{G}$ and its estimator $\widehat{\dyadic{G}}$. In terms of $\dyadic{G}_1$ and $\dyadic{G}_2$,
\begin{align}
\mathcal{D}=&\,\underbrace{\MEAN{}{\Tr{\left(\widehat{\dyadic{G}}_1-\dyadic{G}_1\right)^2}}}_{\mathclap{\displaystyle \qquad\!\!\!=\mathcal{D}_1}}+\underbrace{\MEAN{}{\Tr{\left(\widehat{\dyadic{G}}_2-\dyadic{G}_2\right)^2}}}_{\mathclap{\displaystyle \qquad\!\!\!=\mathcal{D}_2}}\nonumber\\
&\,+\left\{\text{cross terms}\right\}\,.
\label{eq:G1G2MSE}
\end{align}
To illustrate the physics behind moment reconstruction, we shall analyze both the first and second-moment reconstruction accuracy separately. In practice, these analyses are relevant to the situation where the reconstructions of $\dyadic{G}_1$ and $\dyadic{G}_2$ are carried out with independent data. For this situation, the $\left\{\text{cross terms}\right\}$ in Eq.~\eqref{eq:G1G2MSE} vanish so that the total MSE is the sum of the respective MSEs $\mathcal{D}_1$ and $\mathcal{D}_2$ of the reconstructed moments. From hereon, to facilitate discussions, we shall analyze the quantity $\rvec{r}=\left<\rvec{R}\right>$ in place of $\dyadic{G}_1$, where $\mathcal{D}_1=\MEAN{}{(\widehat{\rvec{r}}-\rvec{r})^2}$.

In unbiased statistical estimation theory~\cite{Cox:2006dv}, the MSE $\mathcal{D}\geq\Tr{\dyadic{F}^{-1}}$ is bounded from below by the inverse of the Fisher information matrix $\dyadic{F}$, or the Cram{\'e}r-Rao bound (CRB). Consequently, we have the respective first- and second-moment CRBs $\mathcal{D}_1\geq\Tr{\dyadic{F}_1^{-1}}$ and $\mathcal{D}_2\geq\Tr{\dyadic{F}_2^{-1}}$. Therefore, the general theory of the Fisher matrices $\dyadic{F}_1$ and $\dyadic{F}_2$ for the two CV schemes is in order.

\begin{figure}[t]
	\centering
	\includegraphics[width=1\columnwidth]{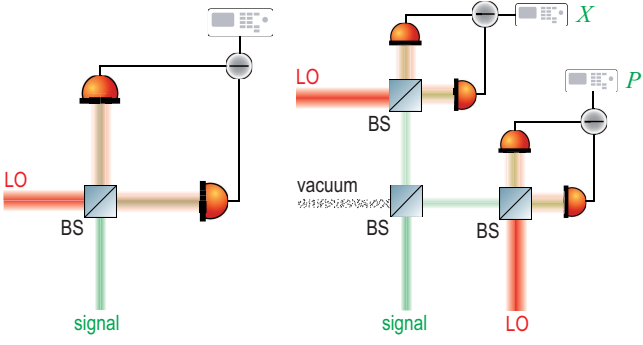}
	\caption{\label{fig:Fig1}Schema for the (a)~homodyne and (b)~heterodyne setups.}
\end{figure}

\section{General theory}
\label{sec:gen_theory}

\subsection{Homodyne detection}

The homodyne detection~\cite{Yuen:1983ba,Abbas:1983ak, Schumaker:1984qm} involves a 50:50 beam splitter that introduces an interference between the optical source of an unknown state (signal) and the local oscillator (coherent-state reference source or simply LO), the latter of which is set to a much larger optical intensity than the mean intensity of the optical source of an unknown quantum state $\rho$~[see Fig.~\ref{fig:Fig1}(a)]. A subtraction of the output photocurrents gives a distribution of voltage readouts $-\infty<x_\vartheta<\infty$ for the LO phase $0\leq\vartheta\leq\pi$, which essentially corresponds to the eigenvalue probability distribution of the quadrature operator $X_\vartheta=X\cos\vartheta+P\sin\vartheta$. It then follows that statistically, the expectation value $\left<X_\vartheta^m\right>$ for any integer value $m$ contains all measurable information about the $m$th operator moments of $X$ and $P$. Since the data acquired with this scheme are the marginals of the Wigner function, the first $(m=1)$ and second $(m=2)$ moments, or $\dyadic{G}$, that are reconstructed with these data may be attributed to this quasi-probability distribution function.

In a typical homodyne experiment, the value of $\vartheta$ is set to increase linearly. The data collected would then be binned for all the measured $\vartheta$ values. The data bins are mutually independent, so that the Fisher matrices $\dyadic{F}_{1,\textsc{hom}}$ and $\dyadic{F}_{2,\textsc{hom}}$ for the respective first- and second-moment CRBs can each be understood as a summation of Fisher matrices of every LO phase bin according to the additivity property of the Fisher information. In the limit of large number of sampling events $N$, the central limit theorem states that the unbiased estimator $\widehat{\left<X^m_\vartheta\right>}$ of the $m$th quadrature moment $\left<X^m_\vartheta\right>$ that is defined as an average sum of independently collected random voltage values for the phase $\vartheta$ follows a Gaussian distribution of data mean $\mu=\mu(\vartheta)=\left<X^m_\vartheta\right>$ and data variance $\sigma^2/N$ where $\sigma^2=\sigma(\vartheta)^2=\left<X^{2m}_\vartheta\right>-\left<X^m_\vartheta\right>^2$, so that the Fisher matrix
\begin{equation}
\dyadic{F}_{\vartheta,m}=\dfrac{N}{\sigma^2}\dfrac{\partial\mu}{\partial\rvec{a}}\dfrac{\partial\mu}{\partial\rvec{a}}+\dfrac{1}{2\sigma^4}\dfrac{\partial\sigma^2}{\partial\rvec{a}}\dfrac{\partial\sigma^2}{\partial\rvec{a}}
\label{eq:fisher_theta}
\end{equation}
for a given LO phase $\vartheta$ in the large-$N$ limit follows the well-known expression for Gaussian distributions, where in our case $\rvec{a}$ is the column of $m$th moment parameters we are interested in reconstructing. As it is clear that only the first term of \eqref{eq:fisher_theta} would survive in this limit, we thus have the \emph{scaled} homodyne Fisher matrix
\begin{equation}
\widetilde{\dyadic{F}}_{m,\textsc{hom}}=\int_{(\pi)}\dfrac{\D\vartheta}{\pi}\,\dfrac{\dyadic{F}_{\vartheta,m}}{N}=\int_{(\pi)}\dfrac{\D\vartheta}{\pi}\,\dfrac{1}{\sigma(\vartheta)^2}\dfrac{\partial\mu(\vartheta)}{\partial\rvec{a}}\dfrac{\partial\mu(\vartheta)}{\partial\rvec{a}}
\label{eq:Fhomm}
\end{equation}
with respect to the number of sampling events $N$ for the complete set of homodyne quadrature-eigenstate outcomes. Scaled statistical quantities such as this one shall be the focus of this article in analyzing tomographic performances, as the scaled CRB (sCRB) represents the power-law coefficient of the MSE in this limit that determines the difficulty in obtaining an estimator of a certain pre-chosen MSE accuracy.

\subsubsection{First-moment reconstruction}

All information about the first moments, $\rvec{a}=\rvec{r}$, of the covariance matrix is completely encoded in the expectation value $\mu(\vartheta)=\langle X_\vartheta\rangle$. The variance for the data is then given by $\sigma(\vartheta)^2=\langle X_\vartheta^2\rangle-\langle X_\vartheta\rangle^2$. The scaled Fisher matrix for the first-moment estimation with homodyne data is therefore given by
\begin{equation}
\FHOMoneN=\int_{(\pi)}\dfrac{\D\theta}{\pi}\dfrac{\dyadic{m}_\vartheta}{\langle X^2_\vartheta\rangle-\langle X_\vartheta\rangle^2}\,,
\label{eq:Fisher_HOM_first}
\end{equation}
where
\begin{equation}
\dyadic{m}_\vartheta\,=\begin{pmatrix}
(\cos\vartheta)^2 & \sin\vartheta\cos\vartheta\\
\sin\vartheta\cos\vartheta & (\sin\vartheta)^2
\end{pmatrix}\,.
\end{equation}
The integral can be evaluated exactly, bringing us to the closed-form expression
\begin{equation}
\mathcal{H}_\textsc{1,hom}=\Tr{\dyadic{G}}+2\sqrt{\DET{\dyadic{G}}}\,.
\label{eq:CRB_HOM_first}
\end{equation}

With the machinery of quantum tomography (see Appendix~\ref{subsec:app:firstmom-est}), an observer can construct the optimal moment estimator that achieves the sCRB. Suppose that the observer collects homodyne data for $N$ sampling events and bins the voltage values into $\{x_{jk}\}$ according to a discrete number $n_\vartheta$ of LO phase bins $\vartheta_k$, where $j$ labels the $n_x$ real voltage values per LO phase bin $\vartheta=\vartheta_k$ and $k$ labels the phase bins. Then an unbiased estimator for any particular expectation value $\langle X_k\rangle\equiv\langle X_{\vartheta_k}\rangle$ would be
\begin{equation}
\widehat{\langle X_k\rangle}=\dfrac{1}{N}\sum^{n_x}_{j=1}n_{jk}x_{jk}\,,\quad \sum^{n_\vartheta}_{k=1}\sum^{n_x}_{j=1}n_{jk}=\sum^{n_\vartheta}_{k=1}N_k=N\,.
\end{equation}
Then upon denoting $\rvec{u}_k\equiv\rvec{u}_{\vartheta_k}=(\cos{\vartheta_k}\,\,\sin{\vartheta_k})^{\textsc{t}}$ where we note that $\dyadic{m}_k=\rvec{u}_k\TP{\rvec{u}}_k$, the optimal first-moment estimator is given by
\begin{align}
\widehat{\rvec{r}}^{(\textsc{opt})}_\textsc{hom}&=\dyadic{W}_1^{-1}\sum^{n_\vartheta}_{k=1}\rvec{u}_k\dfrac{N_k\widehat{\left<X_k\right>}}{\widehat{\left<X^2_k\right>}-\widehat{\left<X_k\right>}^2}\,\nonumber\\
\dyadic{W}_1&=\sum^{n_\vartheta}_{k=1}\rvec{m}_{k}\dfrac{N_k}{\widehat{\left<X_k^2\right>}-\widehat{\left<X_k\right>}^2}\,,
\end{align}
which is immediately computable given the processed data $\left\{\widehat{\left<X_k\right>}\right\}$ and $\left\{\widehat{\left<X^2_k\right>}\right\}$ that are defined by
\begin{equation}
\widehat{\left<X^m_k\right>}=\dfrac{1}{N}\sum^{n_x}_{j=1}n_{jk}x^m_{jk}\quad(m=1,2,\ldots)\,.
\label{eq:data_l}
\end{equation}
That this estimator achieves the sCRB asymptotically is also shown in  Appendix~\ref{subsec:app:firstmom-est}. This equivalently implies that the optimal estimator is as efficient as the maximum-likelihood~(ML) estimator for the multinomially-distributed binned data $\{x_{jk}\}$.

\subsubsection{Second-moment reconstruction}

To estimate $\dyadic{G}_2$, it is clear that second-moment information is completely encoded in the second quadrature moment $\langle X^2_\vartheta\rangle$, which is a function of the three independent parameters $a_1=\langle X^2\rangle$, $a_2=\langle \frac{1}{2}\{\Delta X,\Delta P\}\rangle$ and $a_3=\langle P^2\rangle$. From Eq.~\eqref{eq:Fhomm}, the corresponding $3\times3$ Fisher matrix for these three parameters is
\begin{equation}
\FHOMtwoN=N\int_{(\pi)}\dfrac{\D\theta}{\pi}\dfrac{\dyadic{M}_\vartheta}{\langle X^4_\vartheta\rangle-\langle X^2_\vartheta\rangle^2}\,,
\label{eq:Fisher_HOM}
\end{equation}
where
\begin{equation}
\dyadic{M}_\vartheta\,\,\widehat{=}\begin{pmatrix}
\left(\cos{\vartheta}\right)^2\\
\sqrt{2}\sin{\vartheta}\cos{\vartheta}\\
\left(\sin{\vartheta}\right)^2
\end{pmatrix}\begin{pmatrix}
\left(\cos{\vartheta}\right)^2 & \sqrt{2}\sin{\vartheta}\cos{\vartheta} & \left(\sin{\vartheta}\right)^2
\end{pmatrix}\,.
\end{equation}

The analytical answer to $\FHOMtwoN$ for an arbitrary state, and its subsequent inverse $\mathcal{H}_{2,\textsc{hom}}=\Tr{\FHOMtwoN^{-1}}$ is difficult to calculate, as the denominator in the integrand generally contains trigonometric functions in a complicated manner. Nevertheless, the integral can be calculated explicitly for many interesting and important quantum sources.

The optimal second-moment estimator (see Appendix~\ref{subsec:app:secmom-est}) that achieves the corresponding sCRB can be cleanly expressed using the vectorization operation $\VEC{\dyadic{Y}}$ that turns a matrix into a column according to
\begin{equation}
\dyadic{Y}\,\,\widehat{=}\begin{pmatrix}
y_1 & y_2\\
y_2 & y_3
\end{pmatrix}\quad\mapsto\quad\VEC{\dyadic{Y}}\,\,\widehat{\equiv}\begin{pmatrix}
y_1\\
\sqrt{2}\,y_2\\
y_3
\end{pmatrix}
\end{equation}
in any pre-chosen basis, such that $\Tr{\dyadic{Y}_1\dyadic{Y}_2}=\TP{\VEC{\dyadic{Y}_1}}\VEC{\dyadic{Y}_2}$ for any two $2\times2$ symmetric matrices $\dyadic{Y}_1$ and $\dyadic{Y}_2$. Given the processed data defined in Eq.~\eqref{eq:data_l}, the final operationally-ready expressions for this optimal estimator are given as follows:
\begin{align}
\widehat{\dyadic{G}}^{(\textsc{opt})}_{2,\textsc{hom}}&=\dyadic{W}_2^{-1}\sum^{n_\vartheta}_{k=1}\VEC{\dyadic{m}_k}\dfrac{N_k\widehat{\left<X_k^2\right>}}{\widehat{\left<X^4_k\right>}-\widehat{\left<X_k^2\right>}^2}\,,\nonumber\\
\dyadic{W}_2&=\sum^{n_\vartheta}_{k=1}\dyadic{M}_k\dfrac{N_k}{\widehat{\left<X_{k}^4\right>}-\widehat{\left<X_{k}^2\right>}^2}\,.
\end{align}

For accurate tomography, the value of $N$ is typically large enough such that $\widehat{\dyadic{G}}^{(\textsc{opt})}_{2,\textsc{hom}}$ is a proper covariance matrix and approaches the ML estimator that asymptotically achieves the sCRB, which is strictly speaking the correct regime where $\widehat{\dyadic{G}}^{(\textsc{opt})}_{2,\textsc{hom}}$ is to be used for second-moment tomography. On a separate note, optimal estimators for overcomplete quantum-state tomography of $\rho$ was developed in \cite{Zhu:2014aa} and later rederived in \cite{Teo:2015qs} with the variational principle that is also used to construct the optimal moment estimators in Appendix~\ref{app:optimality}.

\subsection{Heterodyne detection}
\label{subsec:heterodyne}

The heterodyne detection scheme essentially uses two homodyne setups to perform a joint measurement of two complementary observables~[see Fig.~\ref{fig:Fig1}(b)], which are in this case chosen to be the standard $X$ and $P$ quadrature pair for convenience. It is well-known (\cite{Arthurs:1965al,Yuen:1982hh,Arthurs:1988aa,Martens:1990al,Martens:1991aa,Raymer:1994aj,Trifonov:2001up,Werner:2004as}) that the product of their joint-measurement standard deviations has a larger lower bound than the usual one-half of a quantum unit given by the original Heisenberg relation owing to the additional quantum noise introduced by the joint measurement. 

The outcomes for this scheme are in fact the overcomplete set of coherent states. This means that the resulting data are direct phase-space samples of the Husimi~function for the statistical operator $\rho$. The technical complication of having additional measurement noise can therefore be translated completely into the phase-space language that is relevant in our subsequent analysis. Given an infinite set of the Husimi-function data, we have access to the moments $\overline{x^kp^l}$ (the overline denotes the average with respect to the Husimi function, or simply the Husimi average), with which the corresponding ``$\dyadic{G}$'' operator
\begin{equation}
\dyadic{G}_{\textsc{het}}\,\widehat{=}\begin{pmatrix}
\overline{x^2}-\overline{x}^2 & \overline{xp}-\overline{x}\,\overline{p}\\
\overline{xp}-\overline{x}\,\overline{p} & \overline{p^2}-\overline{p}^2
\end{pmatrix}
\end{equation}
can be directly constructed. One can then show that for any quantum state,
\begin{equation}
\dyadic{G}_{\textsc{het}}=\dyadic{G}+\dfrac{\dyadic{1}}{2}\,.
\label{eq:Ghet_G_half}
\end{equation}
The corresponding Arthurs-Kelly type measurement uncertainty relation
\begin{equation}
\VAR{\textsc{q}}{x}\VAR{\textsc{q}}{p}=\left(\langle(\Delta X)^2\rangle+\frac{1}{2}\right)\left(\langle(\Delta P)^2\rangle+\frac{1}{2}\right)\geq1\,,
\end{equation}
which is saturated by coherent states $[\langle(\Delta X)^2\rangle=\langle(\Delta P)^2\rangle=1/2]$, can thereafter be understood as a physical manifestation of the Gauss-Weierstrass transform [related to Eq.~\eqref{eq:Ghet_G_half}] between the Wigner and Husimi functions if the joint-measurement data are directly used to calculate variances (here denoted by $\VAR{\textsc{q}}{y}=\overline{y^2}-\overline{y}^2$ for a complete Husimi-function data $\{y\}$). We shall show that this additional quantum noise, when combined with optimal tomography strategies, can still lead to better moment-reconstruction accuracies relative to the homodyne scheme.

\subsubsection{First-moment reconstruction}

From Sec.~\ref{subsec:heterodyne}, we note that the data collected from the heterodyne scheme are a scatter set of phase-space coordinates $\{(x_j,p_j)\}$ that are distributed according to the Husimi function. As Eq.~\eqref{eq:Ghet_G_half} tells us that there is no difference between the state average $\rvec{r}$ and Husimi average of $(x\,\,\,p)^\textsc{t}$, being a two-parameter estimation scheme, the first-moment sCRB with respect to the heterodyne data can again be found by taking the average of the distance between the estimator
\begin{equation}
\widehat{\rvec{r}}_\textsc{het}\,\widehat{=}\,\dfrac{1}{N}\sum^N_{j=1}\begin{pmatrix}
x_j\\
p_j
\end{pmatrix}
\end{equation}
and the true column $\rvec{r}^\textsc{t}\widehat{=}\,(\overline{x}\,\,\,\overline{p})$:
\begin{equation}
\mathcal{D}_{1,\textsc{het}}=\MEAN{}{\left(\widehat{\rvec{r}}_\textsc{het}-\rvec{r}\right)^2}=\dfrac{1}{N}\left(\VAR{\textsc{q}}{x}+\VAR{\textsc{q}}{p}\right)\,,
\end{equation}
so that
\begin{equation}
\mathcal{H}_{1,\textsc{het}}=\VAR{\textsc{q}}{x}+\VAR{\textsc{q}}{p}=\Tr{\dyadic{G}}+1\,.
\label{eq:CRB_HET_first}
\end{equation}

That $N\mathcal{D}_{1,\textsc{het}}=\mathcal{H}_{1,\textsc{het}}$ follows in the limit of large $N$, where the unbiased estimator $\widehat{\rvec{r}}_\textsc{het}$ is asymptotically optimal since in this limit, the distribution of $\widehat{\rvec{r}}_\textsc{het}$ follows a bivariate Gaussian distribution with vanishing widths, such that $\widehat{\rvec{r}}_\textsc{het}$ becomes the ML estimator that approaches the sCRB for this Gaussian distribution.

\subsubsection{Second-moment reconstruction}

Similarly, to arrive at the optimal accuracy for estimating $\dyadic{G}_2$ using heterodyne data, we define the natural second-moment estimator
\begin{equation}
\widehat{\dyadic{G}}_{2,\textsc{het}}\,\widehat{=}\dfrac{1}{N}\sum^{N}_{j=1}\begin{pmatrix}
x_j^2 & x_jp_j\\
x_jp_j & p_j^2
\end{pmatrix}\,,
\end{equation}
where $\{(x_j,p_j)\}$ are again the sampled Husimi-function data collected during heterodyne detection. From Eq.~\eqref{eq:Ghet_G_half}, we get
\begin{equation}
\dyadic{G}_{2,\textsc{het}}=\dyadic{G}_{2}+\dfrac{\dyadic{1}}{2}\,.
\label{eq:onehalf}
\end{equation}

The MSE $\mathcal{D}_{2,\textsc{het}}$ for heterodyne detection concerning second-moment estimation is consequently given by
\begin{align}
\mathcal{D}_{2,\textsc{het}}&=\MEAN{}{\Tr{\left(\widehat{\dyadic{G}}_{2,\textsc{het}}-\dyadic{G}_{2,\textsc{het}}\right)^2}}\nonumber\\
&=\Tr{\MEAN{}{\widehat{\dyadic{G}}_{2,\textsc{het}}^2}}-\Tr{\dyadic{G}_{2,\textsc{het}}^2}\nonumber\\
&=\dfrac{1}{N}\left(\VAR{\textsc{q}}{x^2}+\VAR{\textsc{q}}{p^2}+2\,\VAR{\textsc{q}}{xp}\right)\,.
\end{align}
In the large-$N$ limit, this MSE is essentially the sCRB 
\begin{equation}
\mathcal{H}_{2,\textsc{het}}=\VAR{\textsc{q}}{x^2}+\VAR{\textsc{q}}{p^2}+2\,\VAR{\textsc{q}}{xp}\,.
\label{eq:CRB_HET}
\end{equation}
since $\widehat{\dyadic{G}}_{2,\textsc{het}}$ again becomes the ML estimator. To see this, we inspect the Fisher matrix $\dyadic{F}_{2,\textsc{het}}$ for the estimator $\widehat{\dyadic{G}}_{2,\textsc{het}}$. If we look at the random column
\begin{equation}
\rvec{x}=\VEC{\widehat{\dyadic{G}}_{2,\textsc{het}}}\widehat{\equiv}\,\dfrac{1}{N}\sum^N_{j=1}\begin{pmatrix}
x_j\\
\sqrt{2}\,x_jp_j\\
p_j
\end{pmatrix}
\end{equation}
that represents $\widehat{\dyadic{G}}_{2,\textsc{het}}$, we find that in the limit of large $N$, the central limit theorem again says that $\rvec{x}$ follows a Gaussian distribution defined by the mean $\rvec{\mu}=\overline{\rvec{x}}\,\widehat{=}\,\TP{(\overline{x^2}\,\,\,\sqrt{2}\,\overline{xp}\,\,\,\,\overline{p^2})}$ and the covariance matrix 
\begin{equation}
\dyadic{\Sigma}=\overline{\rvec{x}\TP{\rvec{x}}}-\rvec{\mu}\TP{\rvec{\mu}}\,\widehat{=}\dfrac{1}{N}\begin{pmatrix}
\VAR{\textsc{q}}{x^2} & * & *\\
* & 2\,\VAR{\textsc{q}}{xp} & *\\
* & * & \VAR{\textsc{q}}{p^2}
\end{pmatrix}\,,
\end{equation}
so that we eventually recover the well-known result $\dyadic{\Sigma}=\dyadic{F}_{2,\textsc{het}}^{-1}$ for Gaussian scatter data that saturates the CRB as we remember that $\Tr{\dyadic{\Sigma}}=\mathcal{D}_{2,\textsc{het}}$. Equation~\eqref{eq:CRB_HET} then follows tout de suite.

\section{First-moment estimation}
\label{sec:first-mom}

\subsection{General optimality of heterodyne tomography}

As far as first-moment estimation is concerned, the general results in Eqs.~\eqref{eq:CRB_HOM_first} and \eqref{eq:CRB_HET_first} imply that $\mathcal{H}_{1,\textsc{het}}\leq\mathcal{H}_{1,\textsc{hom}}$ for \emph{any} quantum state. This main result hinges on the physical HRS uncertainty relation, which is equivalent to the constraint $\DET{\dyadic{G}}\geq1/4$ for the covariance matrix $\dyadic{G}$. This constraint means that
\begin{equation}
\mathcal{H}_{1,\textsc{hom}}=\Tr{\dyadic{G}}+2\sqrt{\DET{\dyadic{G}}}\geq\Tr{\dyadic{G}}+1=\mathcal{H}_{1,\textsc{het}}\,.
\end{equation}
This implies that for \emph{all} quantum states, the reconstruction accuracy of the optimal heterodyne first-moment estimator is always higher or equal to that of the optimal homodyne first-moment estimator in locating the average center of the quantum state in phase space. For minimum-uncertainty states, the accuracies of the two schemes are equal ($\mathcal{H}_{1,\textsc{hom}}=\mathcal{H}_{1,\textsc{het}}$). Subsequent well-known and interesting examples merely illustrate this fundamental fact. In terms of the first-moment performance ratio 
\begin{equation}
\gamma_1=\dfrac{\mathcal{H}_{1,\textsc{het}}}{\mathcal{H}_{1,\textsc{hom}}},
\end{equation}
a subunit magnitude indicates that the heterodyne scheme outperforms the homodyne scheme.

\subsection{Gaussian states}
\label{subsec:mom1_gauss}

For a Gaussian state where the covariance matrix $\dyadic{G}$ characterizes the spread of its Wigner function, the state variance of $X_\vartheta$ is simply
\begin{equation}
\langle X^2_\vartheta\rangle-\langle X_\vartheta\rangle^2=\TP{\rvec{u}}_\vartheta\,\dyadic{G}\,\rvec{u}_\vartheta\,.
\end{equation}
From Eqs.~\eqref{eq:CRB_HOM_first} and \eqref{eq:CRB_HET_first}, the first-moment performance ratio
\begin{equation}
\gamma_1=\dfrac{\Tr{\dyadic{G}}+1}{\Tr{\dyadic{G}}+2\sqrt{\DET{\dyadic{G}}}}\leq1
\end{equation}
clearly cannot exceed one since any physical state satisfying the HRS uncertainty relation must take $\DET{\dyadic{G}}\geq1/4$. The maximum value of $\gamma_1=1$ is attained for minimum-uncertainty states.

\subsection{Fock states}

A Fock state of the ket $\ket{n}$ is always centered at the origin of the phase space $(\rvec{r}=\rvec{0})$. The circular symmetry of these states imply the fact that $(\Delta X)^2=(\Delta P)^2=n+1/2=(\Delta X_\vartheta)^2$, whence
\begin{equation}
\mathcal{H}_{1,\textsc{hom}}=2(2n+1)
\label{eq:CRB_HOM_Fock_first}
\end{equation}
since such states have zero first moments. On the other hand, for the heterodyne scheme, we get
\begin{equation}
\mathcal{H}_{1,\textsc{het}}=2(n+1)
\label{eq:CRB_HET_Fock_first}
\end{equation}
by simply using the Husimi~characteristic function from Table~\ref{tab:char} in Appendix~\ref{app:char_func}. Therefore, we get a
\begin{equation}
\gamma_1=\dfrac{n+1}{2n+1}
\label{eq:gamma_FOCK}
\end{equation}
that is always sub-unity unless $n=0$, a result that is again familiar from Sec.~\ref{subsec:mom1_gauss}. In the limit of large photon numbers, the first-moment $\gamma_1$ approaches 1/2.

\subsection{Even/odd coherent states}

Another popular class of non-Gaussian states with interesting phase-space quantum interference features are the even/odd coherent states characterized by the ket $\ket{\pm;\alpha_0}=(\ket{\alpha_0}\pm\ket{-\alpha_0})\mathcal{N}_\pm$ of appropriate normalization constants $\mathcal{N}_\pm=1/\sqrt{2\pm2\,\E{-2|\alpha_0|^2}}$, whose first moments $\rvec{r}$ are all equal to zero. Using the definitions $a=\frac{1}{2}\left[\left<(\Delta X)^2\right>-\left<(\Delta P)^2\right>\right]=\alpha_0^2$ and $b_\pm=\frac{1}{2}\left[\left<(\Delta X)^2\right>+\left<(\Delta P)^2\right>\right]=\alpha_0^2\left[\tanh(\alpha_0^2)\right]^{\pm1}+1/2$,
\begin{equation}
\mathcal{H}_{1,\textsc{hom}}=2\left(b_\pm+\sqrt{b_\pm^2-a^2}\right)\,.
\label{eq:CRB_HOM_padd_first}
\end{equation}
For the heterodyne counterpart, one finds that
\begin{equation}
\mathcal{H}_{1,\textsc{het}}=2\left(b_\pm+\frac{1}{2}\right)\,,
\end{equation}
which contributes to the performance ratio
\begin{equation}
\gamma_1=\dfrac{b_\pm+\frac{1}{2}}{b_\pm+\sqrt{b_\pm^2-a^2}}\,.
\end{equation}

For both types of coherent state superpositions, $\gamma_1\rightarrow1$ as $\alpha_0\rightarrow\infty$. For even coherent states, the performance ratio $\gamma_1=1$ when $\alpha_0=0$ as it should. Otherwise, this ratio is always less than one for any positive $\alpha_0$. There exists a single local minimum of $\gamma_1\approx0.7577$ at $\alpha_0\approx1.715$. For odd coherent states, $\gamma_1<1$ for \emph{all} $\alpha_0$ values, with the minimum value of $\gamma_1=1/3$ at $\alpha_0=0$. For these states, $\gamma_1$ increases monotonically to one as $\alpha_0$ tends to infinity.

\subsection{Displaced Fock states}

Displacement and photon-addition are two important physical procedures that are frequently discussed in quantum physics. The different orders in which these processes are carried out on the vacuum state give output states of a different nature. Displacing an $m$-photon-added vacuum state by a complex amplitude $\alpha_0$ results in displaced Fock states defined by the ket $D(\alpha_0)\ket{m}$ can
be effectively performed using a beam splitter with a high transmissivity and a strong coherent state~\cite{Paris:1996do,Banaszek:1999qn}.

It can be shown easily that the first-moment sCRBs are indeed given by Eqs.~\eqref{eq:CRB_HOM_Fock_first} and \eqref{eq:CRB_HET_Fock_first}, so that the performance ratio is then completely identical to that of the usual central Fock states in Eq.~\eqref{eq:gamma_FOCK}. This reflects the physical fact that the accuracy in estimating the displacement cannot explicitly depend on where the center of the displaced Fock states is when full sets of CV measurement outcomes are considered, as the tomographic coverage of the entire phase space is then complete. This accuracy depends only on the variances, which describe the second-order symmetry and is unaffected at all by the displacement.

\subsection{Photon-added coherent states}

A swap in the order of photon addition and displacement on the vacuum state gives the photon-added coherent state of $m$ added photons and reference amplitude $\alpha_0$ is defined by the ket $\ket{m;\alpha_0}=\mathcal{N}_{m,|\alpha_0|^2}{A^\dagger}^m\ket{\alpha_0}$ with the bosonic annihilation operator $A$, where the normalization constant $\mathcal{N}_{m,|\alpha_0|^2}=\E{|\alpha_0|^2/2}/\sqrt{m!\pFq{1}{1}{m+1}{1}{|\alpha_0|^2}}$ involves the confluent hypergeometric function of the first kind $\pFq{1}{1}{a}{b}{y}$. The integer value $m$ denotes the extent to which the mean photon number
\begin{equation}
\left<A^\dagger A\right>=(m+1)\dfrac{\pFq{1}{1}{m+2}{1}{|\alpha_0|^2}}{\pFq{1}{1}{m+1}{1}{|\alpha_0|^2}}-1\,,
\label{eq:mean_padd}
\end{equation}
which is always larger than $|\alpha_0|^2+m$ whenever $\alpha_0\neq0$, is increased nonlinearly by the operation by ${A^\dagger}^m$ on the reference coherent ket $\ket{\alpha_0}$. This particular class of quantum states is but one of many possible kinds of photon-added states, which are of interest to the quantum community for testing some fundamental statements~\cite{Parigi:2004qc,Kim:2008bc,Zavatta:2009aa}.

For these photon-added coherent states, the second-order symmetry is now affected by the combined action of the displacement and photon addition, so that $\left<(\Delta X)^2\right>$ and $\left<(\Delta P)^2\right>$ are functions of $m$ and $\alpha_0$. These expressions can be straightforwardly computed with the help of the characteristic functions given in Table~\ref{tab:char} in Appendix~\ref{app:char_func}. By defining
\begin{align}
a  &= - \frac{\alpha_0^2(m+1)}{2\, \pFq{1}{1}{m+1}{1}{\alpha_0^2}^2} 
\left[2(m+1)\,\pFq{1}{1}{m+2}{2}{\alpha_0^2}^2  \right . \nonumber \\
& \left . - (m+2)\,\pFq{1}{1}{m+1}{1}{\alpha_0^2}\pFq{1}{1}{m+3}{3}{\alpha_0^2}\right]
\,,\nonumber\\
b & = m + \frac{1}{2}-\frac{\alpha_0^2 m
	\,\pFq{1}{1}{m+1}{2}{\alpha_0^2}}
{\,\pFq{1}{1}{m+1}{1}{\alpha_0^2}^2}
\left[\,\pFq{1}{1}{m+1}{1}{\alpha_0^2} \right . \nonumber \\
& \left . + m \, \pFq{1}{1}{m+1}{2}{\alpha_0^2}\right] \,,
\end{align}
such that $b>a$, the first-moment sCRB for homodyne detection is of the same form as in Eq.~\eqref{eq:CRB_HOM_padd_first}, namely
\begin{equation}
\mathcal{H}_{1,\textsc{hom}}=2\left(b+\sqrt{b^2-a^2}\right)\,.
\end{equation}
The first-moment sCRB for heterodyne detection is given by
\begin{equation}
\mathcal{H}_{1,\textsc{het}}=2\left[a+(m+1)\frac{\pFq{1}{1}{m+2}{2}{\alpha_0^2}}{\pFq{1}{1}{m+1}{1}{\alpha_0^2}}\right]\,.
\end{equation}
Clearly, when $\alpha_0=0$, the answers in Eqs.~\eqref{eq:CRB_HOM_Fock_first} and \eqref{eq:CRB_HET_Fock_first} for an $m$-number Fock state are reproduced exactly. With $m=0$, the respective sCRBs of a value of 2 for all $\alpha_0$s are furthermore consistent with Sec.~\ref{subsec:mom1_gauss}. Otherwise, $\gamma_1$ is always sub-unity, and approaches unity as $\alpha_0\rightarrow\infty$.

\section{Second-moment estimation}
\label{sec:sec-mom}

\subsection{Gaussian states}
\label{subsec:second_mom_gauss}

It seems fitting to commence the discussion of second-moment estimation with the Gaussian state, for it is natural to begin with the generalization of the results that already appeared in Refs.~\cite{Rehacek:2015qp} and \cite{Muller:2016da} to general noncentral Gaussian states $\left(\rvec{r}\neq\rvec{0}\right)$. We suppose that the Gaussian state of the covariance matrix $\dyadic{G}$ is centered at $\rvec{r}=\rvec{r}_0=\TP{(x_0\,\,\,p_0)}$. From Table~\ref{tab:char} in Appendix~\ref{app:char_func}, by defining $\mu_\vartheta=\TP{\rvec{u}}_\vartheta\,\rvec{r}_0$ and $\sigma_\vartheta^2=\TP{\rvec{u}}_\vartheta\,\dyadic{G}\,\rvec{u}_\vartheta$, the variance for the second quadrature moment reads
\begin{equation}
\left<X_\vartheta^4\right>-\left<X_\vartheta^2\right>^2=2\,\sigma_\vartheta^2\left(\sigma_\vartheta^2+2\,\mu_\vartheta^2\right)\,.
\label{eq:GAUSS_var}
\end{equation}
For \emph{central} Gaussian states $(\langle X\rangle=\langle P\rangle=0)$, we have $\langle X^4_\vartheta\rangle=3\langle X^2_\vartheta\rangle^2$ and the scaled Fisher matrix in Eq.~\eqref{eq:Fisher_HOM} turns into the familiar form in \cite{Rehacek:2015qp,Muller:2016da}. For the more general situation, one can repeat the contour-method integration in \cite{Rehacek:2015qp} to calculate the scaled Fisher matrix in Eq.~\eqref{eq:Fisher_HOM}. The answer is given as
\begin{widetext}
\begin{align}
	\FHOMtwoN=\dfrac{- 2}{(c+\I b)(w_3+\I w_2)} \left [  \dfrac{M_{z=0}}{z_{1+}z_{1-}z_{2+}z_{2-}}+\dfrac{M_{z=z_{1-}}}{z_{1-}(z_{1-}-z_{1+})(z_{1-}-z_{2+})(z_{1-}-z_{2-})}+
	\dfrac{M_{z=z_{2-}}}{z_{2-}(z_{2-}-z_{1-})(z_{2-}-z_{1+})(z_{2-}-z_{2+})}\right
	]
\label{eq:Fisher_GAUSS}
\end{align}
together with the definitions
\begin{align}
a&=\dfrac{1}{2}\Tr{\dyadic{G}}\,,\,\,b=\dfrac{1}{2}\left(\dyadic{G}_{11}-\dyadic{G}_{22}\right)\,,\,\,c=\dyadic{G}_{12}\,,\nonumber\\
w_1&=a+\rvec{r}_0^2\,,\!\quad w_2=b+x_0^2-p_0^2\,,\,\quad w_3=c+2x_0p_0\,,\nonumber\\
z_{1\pm}&=\dfrac{-a\pm\I\sqrt{-a^2+b^2+c^2}}{b-\I c}\,,\,\,z_{2\pm}=\dfrac{-w_1\pm\I\sqrt{-w_1^2+w_2^2+2_3^2}}{w_2-\I w_3}\,,\nonumber\\
M_z&\,\,\widehat{=}\,\dfrac{1}{16}\begin{pmatrix}
(z+1)^4 & -\I\sqrt{2}(z-1)(z+1)^3 & -(z^2-1)^2\\
-\I\sqrt{2}(z-1)(z+1)^3 & -2(z^2-1)^2 & \I\sqrt{2}(z+1)(z-1)^3\\
-(z^2-1)^2 & \I\sqrt{2}(z+1)(z-1)^3 & (z-1)^4
\end{pmatrix}\,.
\end{align}
\end{widetext}
When $\rvec{r}_0=0$, we have $w_1=a$, $w_2=b$ and $w_3=c$ and the scaled Fisher matrix $\FHOMtwoN$ reduces to that for the central Gaussian state in \cite{Rehacek:2015qp}. For the general setting, the full expression of $\mathcal{H}_{2,\textsc{hom}}$ is omitted here in this case due to its complexity. On the other hand, the sCRB with the heterodyne scheme for these noncentral Gaussian states can be calculated directly from Eq.~\eqref{eq:CRB_HET} using the characteristic function in Table~\ref{tab:char} and is given by
\begin{align}
\mathcal{H}_{2,\textsc{het}}=2\,\Big(&\Tr{\dyadic{G}_\textsc{het}}^2-\DET{\dyadic{G}_\textsc{het}}\nonumber\\
&+\TP{\rvec{r}_0}\,\dyadic{G}_\textsc{het}\,\rvec{r}_0+\Tr{\dyadic{G}_\textsc{het}}\rvec{r}_0^2\Big)\,,
\label{eq:CRB_HET_GAUSS}
\end{align}where one immediately verifies the counterpart expression in \cite{Rehacek:2015qp} for the central Gaussian states upon setting $\rvec{r}_0=\rvec{0}$.

At this stage, we reassure ourselves the physics of the problem of second-moment tomography by understanding, first, that in the case where tomography is performed on the \emph{full} covariance matrix $\dyadic{G}$ then the sCRB, which is the minimum of the MSE, should not depend on the orientation of the two-dimensional uncertainty region (here being an ellipse for any Gaussian state) described by the eigenvectors of this matrix but only its eigenvalues owing to the form of the MSE. Additionally, the accuracy should also be independent of $\rvec{r}_0$. When only the second-moment matrix $\dyadic{G}_2$ is reconstructed, the sCRB should also not depend on its eigenvectors but only its eigenvalues. The physics remains the same. However, there is a difference between estimating the full matrix $\dyadic{G}$ and estimating just $\dyadic{G}_2$. Since $\dyadic{G}_2$ is in general an increasing function of the first moments, this means that as the displacement of the center from the phase-space origin for the quantum state increases, the geometric mean of eigenvalues (GME) of $\dyadic{G}_2$ correspondingly becomes larger so that the second-order-``temperature'' of the state, a terminology borrowed from Gaussian states, as described by the GME is now higher and this results in a stronger $\dyadic{G}_2$-``thermal'' property much like the thermal Gaussian states. So we would expect, based on the findings in \cite{Rehacek:2015qp}, that states with large displacements give poor second-moment tomographic accuracies for \emph{both} CV schemes, and yet provides a subunit 
\begin{equation}
\gamma_2=\dfrac{\mathcal{H}_{2,\textsc{het}}}{\mathcal{H}_{2,\textsc{hom}}}
\end{equation} 
performance ratio. It is also physically intuitive that the accuracies for both schemes should also be independent of the angle of displacement, but depend only on the magnitude of the displacement. For non-Gaussian states, the \emph{fourth} moments arising from the structure of the MSE, which are no longer functions of the first and second moments as is the case for Gaussian states, also contribute to the sCRB, and therefore $\gamma_2$, as described in the general theory in Sec.~\ref{sec:gen_theory}.

This physics, however, \emph{seems} to be violated by the noncentral-Gaussian-state expressions in \eqref{eq:Fisher_GAUSS} and \eqref{eq:CRB_HET_GAUSS}, namely that $\mathcal{H}_{2,\textsc{het}}$ depends on the explicit displacement vector $\rvec{r}_0$ and covariance matrix $\dyadic{G}$, for instance. This mishap has nothing to do with any kind of physical violation, but has only to do with the way we specify Gaussian states. By choosing to parametrize a multivariate Gaussian distribution using the natural independent parameters $\rvec{r}_0$ and $\dyadic{G}$ (the full matrix), we inadvertently change the eigenvalues of $\dyadic{G}_2$ by changing $\rvec{r}_0$ and fixing $\dyadic{G}$. This becomes obvious when one finds that the two positive eigenvalues $\lambda_\pm$ of $\dyadic{G}_2$ is given by
\begin{equation}
\lambda_{\pm}=|\alpha_0|^2+\dfrac{1}{2}\Tr{\dyadic{G}_\textsc{het}}\pm\left|\alpha_0^2+\TP{\rvec{w}}\,\dyadic{G}_\textsc{het}\,\rvec{w}\right|^2\,,
\label{eq:GAUSS_EIGs}
\end{equation}
where $\rvec{w}= \tfrac{1}{\sqrt{2}} \TP{(1\,\,\,\,\I)}$ and
$\alpha_0=(x_0+\I p_0)/\sqrt{2}$. The consequence of this natural definition results in such an apparent observation. The noncentral Gaussian states so defined form the singular example in this article where this happens, and the two other noncentral non-Gaussian states which we shall soon visit do not have this technical issue.

\begin{figure}[htp]
	\centering
	\includegraphics[width=1\columnwidth]{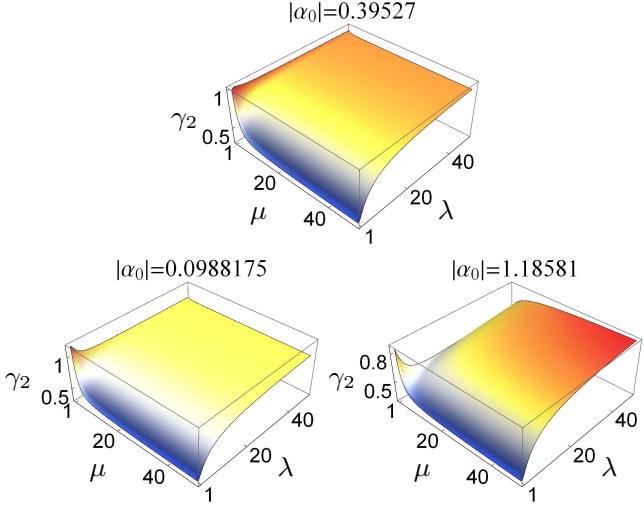}
	\caption{\label{fig:Fig2}Plots of $\gamma_2$ surfaces for $\phi=0$ and different displacement magnitudes along the $x$-axis in phase space. The center plot refers to the critical displacement magnitude of $\sqrt{5/32}\approx0.395$, beyond which $\gamma_2<1$ for all $\mu$ and $\lambda$. The surface tip at $\mu=\lambda=1$ for the coherent states is invariant under a displacement rotation. It is clear from these plots that increasing the temperature reduces the value of $\gamma_2$, while increasing the squeezing strength counters this reduction.}
\end{figure}

To investigate the second-moment performance ratio $\gamma_2=\mathcal{H}_\textsc{2,het}/\mathcal{H}_\textsc{2,hom}$, we may reparametrize the eigenvalues of $\dyadic{G}$ with the squeezing strength $1\leq\lambda<\infty$ and the temperature parameter $1\leq\mu<\infty$ that is commonly adopted in describing all Gaussian states. Then $\dyadic{G}$ has the spectral decomposition
\begin{equation}
\dyadic{G}\,\,\widehat{=}\begin{pmatrix}
\cos\phi & -\sin\phi\\
\sin\phi & \cos\phi
\end{pmatrix}
\begin{pmatrix}
\dfrac{\mu}{2\lambda} & 0\\
0 & \dfrac{\mu\lambda}{2}
\end{pmatrix}
\begin{pmatrix}
\cos\phi & \sin\phi\\
-\sin\phi & \cos\phi
\end{pmatrix}
\label{eq:G_spec}
\end{equation}
where $\phi$ orientates the eigenvectors of $\dyadic{G}$. In this parametrization, we can clearly see that a large displacement magnitude contributes to a large temperature, so that a small value of $\gamma_2$ can be anticipated for these highly displaced or $\dyadic{G}_2$-thermal Gaussian states based on the conclusions in \cite{Rehacek:2015qp} and \cite{Muller:2016da}. The behavior of $\gamma_2$ is very similar to that for the central Gaussian states and is plotted in Fig.~\ref{fig:Fig2} for various values of $|\alpha_0|$. The lowest achievable $\gamma_2$ values go with the highly thermal Gaussian states ($\lambda=1$, $\mu\gg|\rvec{r}_0|$), whose covariance matrix $\dyadic{G}=\mu\dyadic{1}/2$ is simply a multiple of the identity. Their second quadrature moment has a variance $\left<X_\vartheta^4\right>-\left<X_\vartheta^2\right>^2=\mu^2/2$, according to Eq.~\eqref{eq:GAUSS_var}, that is of course independent of the LO phase $\vartheta$ due to the rotational symmetry. The performance ratio then takes the minimum value of $3/10$.

\begin{figure}[htp]
	\centering
	\includegraphics[width=1\columnwidth]{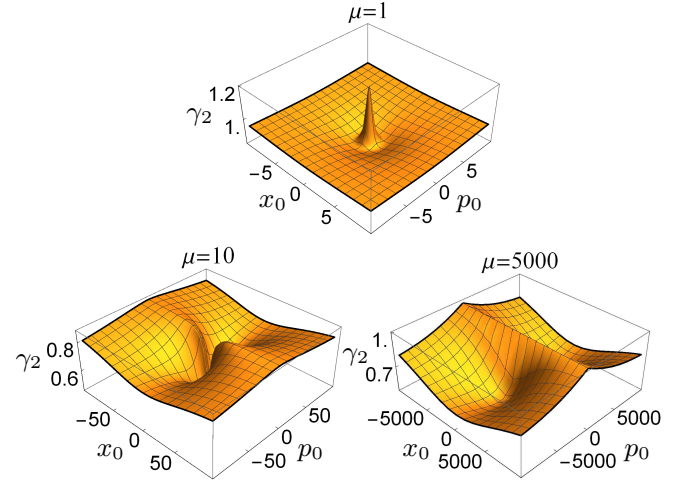}
	\caption{\label{fig:Fig3}Plots of $\gamma_2$ surfaces against the displacement $\rvec{r}_0$ for $\phi=0$ and different values of $\mu=\lambda$. In the limit $\mu\rightarrow\infty$, $\gamma_2\leq1$ approaches unity at $p_0=0$. The signature peak of $\gamma_2=6/5=1.2$ at $x_0=p_0=0$ for $\mu=1$ is consistent with the finding in Refs.~\cite{Rehacek:2015qp} and \cite{Muller:2016da} for central Gaussian states.}
\end{figure}

The maximum of $\gamma_2$ occurs with the coherent states ($\mu=\lambda=1$) and takes a value of $6/5$ for $\rvec{r}_0=\rvec{0}$. For larger magnitudes of $\alpha_0$, the value of $\gamma_2$ drops below unity beyond the magnitude of $|\alpha_0|=\sqrt{5/32}$, which can be obtained through optimization. One may verify that at this critical magnitude, $\mathcal{H}_{2,\textsc{hom}}=\mathcal{H}_{2,\textsc{het}}=63/8$. So, given a displacement magnitude larger than $\sqrt{5/32}$, the heterodyne scheme always outperforms the homodyne scheme in second-moment estimation. In the limit of large $\mu$ and $\lambda$, where we may take this limit such that $\mu=\lambda$ without loss of generality, if one considers the spectral decomposition in Eq.~\eqref{eq:G_spec}, then $\gamma_2$ for $\phi=0$ plotted in Fig.~\ref{fig:Fig3} shows the values for different $\mu$ as an indication that $\gamma_2\leq1$ in this limit. Different $\phi$ values simply rotate these plots in the $x_0$--$p_0$ plane.

\subsection{Fock states}
\label{subsec:second_mom_fock}
Owing to the rotational symmetry of the Fock states [$\dyadic{G}_2=(n+1/2)\dyadic{1}$], the second and fourth quadrature moments
\begin{equation}
\left< X_\vartheta^4\right>-\langle X_\vartheta^2\rangle^2=\dfrac{1}{2}\langle X_\vartheta^2\rangle^2+\dfrac{3}{8}
\label{eq:fock_Xt}
\end{equation}
are independent of the local-oscillator phase $\vartheta$, so that the Fisher matrix
\begin{equation}
\dyadic{F}_{2,\textsc{hom}}=\dfrac{1}{4(n^2+n+1)}\begin{pmatrix}
3 & 0 & 1\\
0 & 2 & 0\\
1 & 0 & 3
\end{pmatrix}\,.
\end{equation}
It then follows that the sCRB is given by
\begin{equation}
\mathcal{H}_{2,\textsc{hom}}=5\,(n^2+n+1)\,.
\label{eq:CRB_HOM_Fock}
\end{equation}
On the other hand, the Husimi characteristic function for the Fock states in Appendix~\ref{app:char_func} produces the answer
\begin{equation}
\mathcal{H}_{2,\textsc{het}}=2\,(n+1)(n+3)\,.
\label{eq:CRB_HET_Fock}
\end{equation}

\begin{figure}[h!]
	\centering
	\includegraphics[width=0.9\columnwidth]{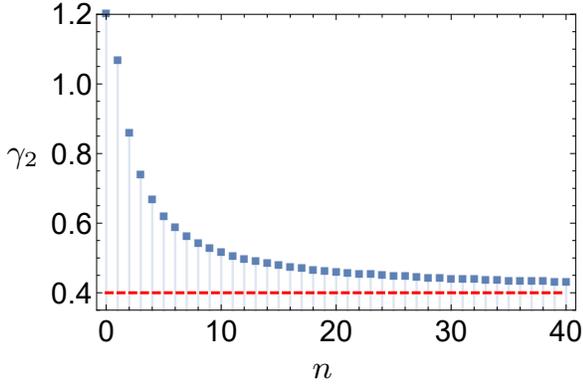}
	\caption{\label{fig:Fig4}Plot of $\gamma_2$ (solid blue circles) against $n$ for Fock states. As $n$ increases, $\gamma_2$ decreases monotonically and eventually saturates at a subunit constant of $2/5$ (dashed red line).}
\end{figure}

The performance ratio 
\begin{equation}
\gamma_2=\dfrac{2\,(n+1)(n+3)}{5\,(n^2+n+1)}
\end{equation}
is less than one for $n\geq2$, in which regime the Fock states are sufficiently $\dyadic{G}_2$-``thermal''. For $n=0$, we evidently obtain the familiar answer $\gamma_2=6/5$ for the vacuum state, whereas for $n=1$, $\gamma_2=16/15$. In the limit of large $n$, $\gamma_2\rightarrow2/5$ (see Fig.~\ref{fig:Fig4}).

\subsection{Even/odd coherent states}
Since the eigenvalues 
\begin{equation}
\lambda^{(\pm)}_\pm=\dfrac{1}{2}+|\alpha_0|^2\left\{\left[\tanh\!\left(|\alpha_0|^2\right)\right]^{(\pm1)}\pm 1\right\}
\end{equation}
of $\dyadic{G}_2$ are simple functions of $|\alpha_0|^2$ for the even/odd $(\pm)$ coherent states, we may take $\alpha_0\geq0$ without loss of generality. The quadrature moments can be easily derived with the help of Appendix~\ref{app:char_func}, which give the following second-moment variance
\begin{align}
\left<X_\vartheta^4\right>-\left<X_\vartheta^2\right>^2=&\,\dfrac{1}{2}+2\alpha_0^2\left\{\cos(2\vartheta)+\left[\tanh\!\left(\alpha_0^2\right)\right]^{\pm1}\right\}\nonumber\\
&\,\pm\dfrac{4\alpha_0^4}{\left(\E{\alpha_0^2}\pm\E{-\alpha_0^2}\right)^2}\,.
\end{align}
By relying on the asymptotic behaviors $\coth y\approx 1/y$ and $\csch y\approx1/y$
of the hyperbolic trigonometric functions for small arguments, we revert to the limiting second-moment variances for $n=0$ and $n=1$, which is consistent with the fact that the even states approach the vacuum state and the odd states approach the single-photon Fock state. The Fisher matrix $\dyadic{F}_{2,\textsc{hom}}$ thus takes the simple form
\begin{align}
\dyadic{F}_{2,\textsc{hom}}&=\int_{(\pi)}\dfrac{\D\vartheta}{\pi}\,\dfrac{\dyadic{M}_\vartheta}{m_\pm+l\cos(2\vartheta)}\quad (l=2\alpha_0^2<m_\pm)\,,\nonumber\\
m_\pm&=\dfrac{1}{2}+2\alpha_0^2\left[\tanh\!\left(\alpha_0^2\right)\right]^{\pm1}\pm\dfrac{4\alpha_0^4}{\left(\E{\alpha_0^2}\pm\E{-\alpha_0^2}\right)^2}\,,
\end{align}
whence one obtains
\begin{equation}
\mathcal{H}_{2,\textsc{hom}}=6m_\pm+4\sqrt{m_\pm^2-l^2}
\label{eq:CRB_HOM_oe}
\end{equation}
after carrying out the integration, matrix inversion and matrix trace. On the other hand, the Husimi-average moments of the heterodyne data contribute to the result
\begin{equation}
\mathcal{H}_{2,\textsc{het}}=6+12\alpha_0^2\left[\tanh\!\left(\alpha_0^2\right)\right]^{\pm1}\pm\dfrac{8\alpha_0^4}{\left(\E{\alpha_0^2}\pm\E{-\alpha_0^2}\right)^2}
\label{eq:CRB_HET_oe}
\end{equation}
for the heterodyne sCRB.

\begin{figure}[h!]
	\centering
	\includegraphics[width=0.9\columnwidth]{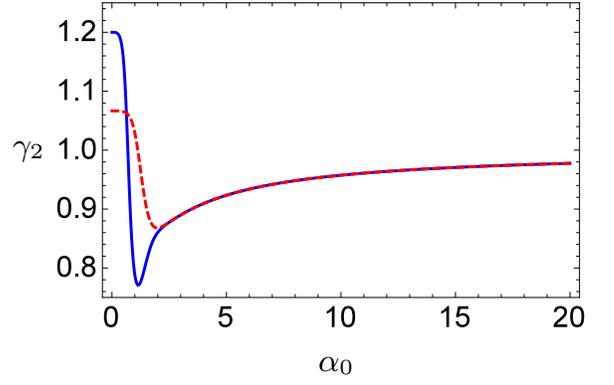}
	\caption{\label{fig:Fig5}Plots of $\gamma_2$ for the even~(solid blue curve) and odd~(dashed red curve) coherent states against the parameter $\alpha_0$ that characterizes the even/odd coherent states. For the even coherent states, the unit-$\gamma_2$ crossover occurs at $\alpha_0\approx0.693$, whereas for the odd coherent states, this happens at $\alpha_0\approx1.128$. Furthermore, for each type of states, $\gamma_2$ possesses a stationary global minimum. For the even states, the minimum value of $\gamma_{2,\text{min}}=0.77096$ is attained at $\alpha_0=1.148\approx1$. For the odd states, this optimum value is $\gamma_{2,\text{min}}=0.86796$ and is achieved with $\alpha_0=1.980\approx2$.}
\end{figure}

We once again remind the reader that the sCRBs stated in Eqs.~\eqref{eq:CRB_HOM_oe} and \eqref{eq:CRB_HET_oe} are independent of the phase of the even/odd coherent states, as this phase amounts to a rotation in phase space that is immaterial in determining the moment-estimation accuracy. For arbitrary complex values of $\alpha_0$, the expressions are still valid after the change $\alpha_0^2\rightarrow|\alpha_0|^2$.

The ratio $\gamma_2$ is greater than one for small values of $\alpha_0$, with the special limiting cases ($\alpha_0=0$) being those of the respective Fock states, and less than one for large values of $\alpha_0$. The crossover values for which these states become sufficiently $\dyadic{G}_2$-``thermal'' such that $\gamma_2=1$ differ for both the even and odd states (see Fig.~\ref{fig:Fig5}). For sufficiently large $\alpha_0$, $\gamma_2$ approaches unity from below. This can be clearly seen by taking the limit $\alpha_0\rightarrow\infty$. In this limit, we have $m_\pm\rightarrow2\alpha_0^2=l$ so that $\mathcal{H}_{2,\textsc{hom}}\rightarrow12\alpha_0^2\approx\mathcal{H}_{2,\textsc{het}}$. For these class of states, $\gamma_2$ has a stationary minimum that is again different for the two types of states, and this is elucidated in Fig.~\ref{fig:Fig5}. At $\alpha_0\approx0.631$, the $\gamma_2$ values for the even and odd states are equal, even though their $\dyadic{G}_2$ matrices are very different. The reason is that the combined contributions of all the second and fourth moments give an overall multiplicative factor of about 2.0694 to both $\mathcal{H}_{2,\textsc{het}}$ and $\mathcal{H}_{2,\textsc{hom}}$ for the odd state relative to the even state.

\subsection{Displaced Fock states}

As opposed to the previous three classes of states, the displaced Fock states (as well as the photon-added coherent states that follow) possess a nonzero quadrature first moment. As the only two parameters $\alpha_0=(x_0+\I p_0)/\sqrt{2}$ and $m$ that characterize these displaced Fock states do not, in any way, restrict the covariance matrix $\dyadic{G}$, it is easy to show that the $\dyadic{G}_2$ geometry, and hence its reconstruction accuracy, depends only on the displacement magnitude $|\alpha_0|^2$ and not its phase. This is done by directly inspecting the eigenvalues of $\dyadic{G}_2$, namely
\begin{align}
\lambda_1&=m+\dfrac{1}{2}\,,\nonumber\\
\lambda_2&=m+2|\alpha_0|^2+\dfrac{1}{2}\,,
\end{align}
one of which is an increasing function of $|\alpha_0|^2$. As a result, we only need to consider the case where $\alpha_0=x_0/\sqrt{2}$ is positive. As $\alpha_0$ increases, the GME increases, which means that the quantum state becomes more $\dyadic{G}_2$-``thermal''. We shall soon see that an increase in $|\alpha_0|^2$ results in a smaller performance ratio $\gamma_2$ in favor of the heterodyne scheme. 

To calculate the homodyne sCRB, we first note that the relevant even-order quadrature moments (see Appendix~\ref{app:char_func}) supply the second-moment quadrature variance
\begin{align}
\left<X_\vartheta^4\right>-\left<X_\vartheta^2\right>^2&=m_0+l\cos(2\vartheta)\,,\nonumber\\
m_0&=\dfrac{1}{2}\left[m^2+m+\alpha_0^2\,(8m+4)\right]\,,\nonumber\\
l&=2\,\alpha_0^2\,(2m+1)<m_0\,,
\end{align}
which bears striking resemblance in form with that for the even/odd coherent states, so that the sCRB also takes the same closed form as Eq.~\eqref{eq:CRB_HOM_oe} inasmuch as
\begin{equation}
\mathcal{H}_{2,\textsc{hom}}=6\,m_0+4\sqrt{m_0^2-l^2}\,.
\label{eq:CRB_HOM_disp}
\end{equation}
For the heterodyne scheme, we subsequently get
\begin{equation}
\mathcal{H}_{2,\textsc{het}}=2\,(m+1)(m+6\,\alpha_0^2)
\label{eq:CRB_HET_disp}
\end{equation}
by again referring to Table~\ref{tab:char}.

The interplay between the discrete ($m$) and continuous $(\alpha_0)$ parameters give rise to familiar cases that have already been analyzed previously for the Gaussian and Fock states. For $m=0$, we of course have the coherent state of amplitude $\alpha_0$ where the maximum $\gamma_2(\alpha_0=0)=6/5$ and the crossover point $\gamma_2(\alpha_0=\sqrt{5/32})=1$ beyond which $\gamma_2<1$ are reproduced by Eqs.~\eqref{eq:CRB_HOM_disp} and \eqref{eq:CRB_HET_disp}. For $m=1$, we have the $m=1$ Fock state for $\alpha_0=0$ so that the unsurprising number $\gamma_2(\alpha_0)=16/15$ comes up from the same sCRB expressions. The crossover point for $\gamma_2=1$ is located at $\alpha_0=\frac{1}{2}\sqrt{19/3-2\sqrt{87}/3}\approx0.1696$. The performance ratio becomes subunit for \emph{all} displacements $\alpha_0$ for $m\geq2$, just like the Fock states. In the limit of large displacements $\alpha_0^2\gg m$, we have $m_0\rightarrow l$ and
\begin{equation}
\gamma_2\!\left(\alpha_0^2\gg m\right)=\dfrac{m+1}{2m+1}\,,
\end{equation}
which approaches $1/2$ in the regime $\alpha_0^2\gg m\gg1$.

\begin{figure}[t]
	\centering
	\includegraphics[width=1\columnwidth]{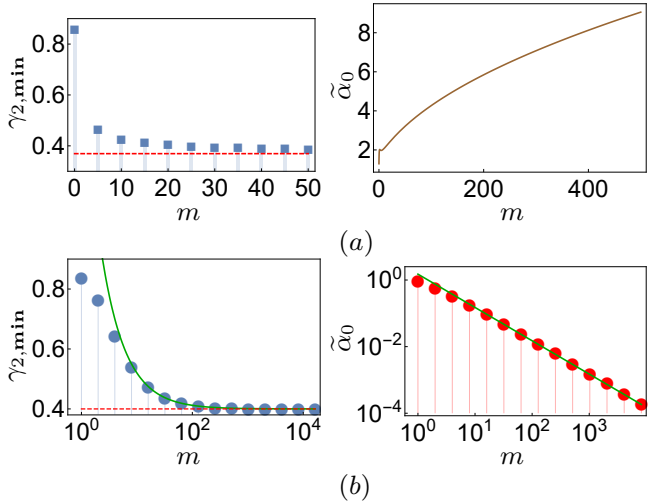}
	\caption{\label{fig:Fig6}Plots of (a)~the optimum (minimum) $\gamma_2$ over all $\alpha_0$ with $m$ (Left) and the minimum point $\alpha_0=\widetilde{\alpha}_0$ (Right) for the displaced Fock states, as well as those of (b) the photon-added coherent states. For the displaced Fock states in (a), $\gamma_{2,\text{min}}$ tends to the limiting value of $6/5-\sqrt{69}/10\approx0.3693$ (dashed red line), and the brown curve representing the exact expression for $\widetilde{\alpha}_0$ shows the quadratic behavior for small $m$ and the approximate square-root behavior for large $m$. On the other hand, for the photon-added states in (b), the numerically found $\gamma_{2,\text{min}}$ values (solid blue circles) are plotted with the theoretical asymptotic power-law curve (solid dark green curve) to illustrate the accuracy of the latter for $m\gtrsim10$, both of which approach the limiting value of $2/5$ (dashed red line). The $\gamma_{2,\text{min}}$ value for $m=0$ (not plotted) has the analytical value of $3(6-\sqrt{21})/5\approx0.85$ that occurs at $\widetilde{\alpha}_0=\sqrt{13+3\sqrt{21}}/4\approx1.29$. The approximate model [see~Eq.~\eqref{eq:model_a0}] for $\widetilde{\alpha}_0$ (green line) is compared with the numerical minima (solid red circles) as a showcase of its remarkable fit.}
\end{figure}

For this two-parameter quantum state, it is interesting to look at the minimum value of $\gamma_2$ over all possible displacement magnitudes $\alpha_0$ for each $m$ [see Fig.~\ref{fig:Fig6}(a)]. To calculate the minimum stationary points $\alpha_0=\widetilde{\alpha}_0$, we differentiate $\gamma_2$ with respect to $\alpha_0$ and set the derivative to zero. While the analytical form for the optimal $\gamma_2=\gamma_{2,\text{min}}$ as a complicated function of $m$ exists, the approximated forms
\begin{equation}
\gamma_{2,\text{min}}\approx\begin{cases}0.8504-0.5893\,m&(\text{small-}m\,\text{ regime})\\
0.3693+\dfrac{0.6565}{m}&(\text{large-}m\,\text{ regime})\end{cases}
\end{equation}
are enough to understand the optimal-$\gamma_2$ curve in terms of a power law already for moderately large $m$. Interestingly, the saturation point for $\gamma_2$ is slightly lower than $2/5$, which is the $\gamma_2$ for the Fock state of an infinitely large $m$ value. This hints that the optimal center for the displaced Fock state of a large $m$ for which $\gamma_2=\gamma_{2,\text{min}}$ is significantly far away from the phase-space origin. This is indeed consistent with the behavior of the minimum point $\widetilde{\alpha}_0$, which also has a complicated closed-form expression~[plotted in Fig.~\ref{fig:Fig6}(a)], so that we only present the more useful approximated forms
\begin{equation}
\widetilde{\alpha}_0\approx\begin{cases}1.2929+2.2060\,m-3.2976\,m^2\!\!\!\!&(\text{small-}m\,\text{ regime})\\
0.3993\sqrt{m}+\dfrac{2.8174}{\sqrt{m}}&(\text{large-}m\,\text{ regime})\end{cases}
\end{equation}
that highlight the main gradient features. To summarize, the minimum value of $\gamma_2$ essentially behaves as a power law in $m$, and the corresponding stationary minimum $\widetilde{\alpha}_0$ is quadratic for small $m$ and goes as a square-root curve for large $m$. 

\subsection{Photon-added coherent states}

As in the case of the displaced Fock states, the eigenvalues of $\dyadic{G}_2$ for the photon-added coherent states,
\begin{align}
\lambda_1=&\,(m+1)\dfrac{\pFq{1}{1}{m+2}{2}{|\alpha_0|^2}}{\pFq{1}{1}{m+1}{1}{|\alpha_0|^2}}-\dfrac{1}{2}\,,\nonumber\\
\lambda_2=&\,2m+2|\alpha_0|^2+\dfrac{1}{2}\nonumber\\
&\,+m(2|\alpha_0|^2-1)\dfrac{\pFq{1}{1}{m+1}{2}{|\alpha_0|^2}}{\pFq{1}{1}{m+1}{1}{|\alpha_0|^2}}\,,
\end{align}
are also functions of $|\alpha_0|^2$, which correctly coincides with the physics of the second-moment estimation problem. This also means that discussing in terms of the range $\alpha_0\geq0$ covers the tomography analysis sufficiently. Moreover, the eigenvalues are increasing functions of the displacement magnitude, so that the GME becomes larger with $\alpha_0$, thereby rendering the photon-added states more $\dyadic{G}_2$-``thermal''. This again gives a smaller performance ratio $\gamma_2$, or a better tomographic performance for the heterodyne scheme compared to the homodyne scheme. 

Once more with the help of Table~\ref{tab:char} in Appendix~\ref{app:char_func}, the quadrature moments can be written down in principle, but they are represented by bulky expressions that are hardly worth any analytical value and the Fisher-matrix integral in Eq.~\eqref{eq:Fisher_HOM} has no known closed-form expression. However, we may still briefly discuss the important limiting cases. For $\alpha_0\ll\sqrt{m}$, to second order in $\alpha_0$, it can be shown that
\begin{equation}
\mathcal{H}_{2,\textsc{hom}}\approx5(m^2+m+1)+10\alpha_0^2(m+1)(m+2)\,,
\end{equation}
where the asymptotic connection with Fock states is clear. On the other hand, in the regime of large $\alpha_0\gg\sqrt{m}$, we find that
\begin{equation}
\mathcal{H}_{2,\textsc{hom}}=3+12\alpha_0^2+2\sqrt{1+8\alpha_0^2}\approx12\alpha_0^2\,,
\label{eq:CRB_HOM_padd_large}
\end{equation}
which is the second-moment homodyne sCRB for coherent states. This is also the homodyne sCRB for large-intensity even/odd coherent states. The reason is that for large amplitudes, all these states behave like a coherent state of amplitude $\alpha_0$ as far as second-moment estimation is concerned since all their $\dyadic{G}_2$ eigenvalues are indistinguishable in this limit.

Upon revisiting Eq.~\eqref{eq:CRB_HET}, the heterodyne sCRB can be shown to have the closed form
\begin{align}
\mathcal{H}_{2,\textsc{het}}=\,2\Bigg\{&3+4m+2\alpha_0^2(m+3)-m\dfrac{\pFq{1}{1}{m+1}{2}{\alpha_0^2}}{\left[\pFq{1}{1}{m+1}{1}{\alpha_0^2}\right]^2}\nonumber\\
&\times\Big[2(\alpha_0^4-3\alpha_0^2-m)\,\pFq{1}{1}{m+1}{1}{\alpha_0^2}\nonumber\\
&\quad\,\,\,\,+m(2\alpha_0^4-2\alpha_0^2+1)\,\pFq{1}{1}{m+1}{2}{\alpha_0^2}\Big]\Bigg\}\,.
\end{align}for $\alpha_0>0$.
The behavior to leading order in $\alpha_0^2$ for $\alpha_0\ll\sqrt{m}$,
\begin{equation}
\mathcal{H}_{2,\textsc{het}}\approx\left(2+4\alpha_0^2\right)(m+1)(m+3)\,,
\label{eq:CRB_HET_padd_small}
\end{equation}
is evidently consistent with the known result for Fock states. For $\alpha_0\gg\sqrt{m}$, we once again have $\mathcal{H}_{2,\textsc{het}}\approx12\alpha_0^2$. Note also that, as expected, the first equality in Eq.~\eqref{eq:CRB_HOM_padd_large} is equal to 5, and Eq.~\eqref{eq:CRB_HOM_padd_large} gives a value of 6 for the vacuum state ($m=0$).

For $m\geq2$, the ratio $\gamma_2<1$ for all $\alpha_0$. This natural extension to the result for the Fock states means that for highly nonlinear photon-``adding'' operations, the performance of heterodyne detection is always better than that of homodyne detection in terms of second-moment covariance-dyadic estimation. For $m=0$, the analysis reverts to that for the coherent state, where the crossover occurs at $\alpha_0=\sqrt{5/32}$ after solving for $\mathcal{H}_{2,\textsc{hom}}=\mathcal{H}_{2,\textsc{het}}=2\left(3+6\alpha_0^2\right)$ so that $\gamma_2(\alpha_0>\sqrt{5/32})<1$, which is again consistent with Sec.~\ref{subsec:second_mom_gauss}. For $m=1$, the crossover point $\alpha_0\approx0.2$ may be obtained as the numerical solution. As $\alpha_0$ approaches infinity, previous arguments imply that $\gamma_2\rightarrow1$ for any $m$.

In view of the behavior of $\gamma_2$, another interesting limit is the high-nonlinearity limit ($m\rightarrow\infty$). In this case, we notice that the value $\alpha_0=\widetilde{\alpha}_0$ for which $\gamma_2$ is minimum approaches zero. A good model to estimate this minimum point in this limit is given by
\begin{equation}
\widetilde{\alpha}_0\approx\dfrac{3}{2m}\,,\label{eq:model_a0}
\end{equation}
which can be approximated from curve fitting. Therefore in the large-$m$ limit, the optimum performance ratio $\gamma_2$ is that of an intense Fock state of a large photon number and so we expect the minimum value of $\gamma_2$ to approach $2/5$ as discussed in Sec.~\ref{subsec:second_mom_fock}. In other words, for sufficiently large $m$, the minimum of $\gamma_2$ follows the noncentral power law
\begin{equation}
\min_{\alpha_0}\{\gamma_2\}=\gamma_2\Big|_{\alpha_0=\widetilde{\alpha}_0}\approx\dfrac{2}{5} + \dfrac{6}{5m}\,.
\end{equation}
Figure~\ref{fig:Fig6}(b) succinctly highlights these observations.

\section{Conclusion}
\label{sec:conc}

We compare the moment-reconstruction performances of the homodyne and heterodyne joint-measurement measurement schemes using optimal moment estimators that minimizes the mean squared-error. We first showed that in first-moment tomography, the heterodyne scheme is always tomographically superior to, or at least as good as, the homodyne scheme for \emph{all} quantum states in terms of the mean squared error of the moment estimators. The underlying physical reason is solely the Heisenberg-Robertson-Schr{\"o}dinger uncertainty relation for complementary observables. For second-moment tomography, we showed that the heterodyne scheme can often outperform the homodyne scheme for Gaussian states and many other interesting and important classes of non-Gaussian states. All these states indicate a trend that a larger geometric mean of second-moment eigenvalues (second-moment ``temperature'') improves the moment reconstruction accuracy with the heterodyne scheme relative to the homodyne scheme. This trend, however, is not monotonic in the second-moment ``temperature'', because there is also influence from the fourth moments originating from the form of the mean squared-error, the combined contributions of both give interesting features the reconstruction accuracy, as illustrated by the examples in this article. The general theory introduced in Sec.~\ref{sec:gen_theory} can be applied to higher-moment estimation that are important in general operator-moment applications and source-calibration protocols, and these shall be reported in the future.

\section{Acknowledgments}
We acknowledge
financial support from the BK21 Plus Program (21A20131111123) funded by the Ministry of Education (MOE, Korea) and National Research Foundation of Korea (NRF), the NRF grant funded by the Korea government (MSIP) (Grant No. 2010-0018295), the Korea Institute of Science and Technology Institutional Program (Project No. 2E26680-16-P025), the European Research Council (Advanced Grant PACART), the Spanish MINECO (Grant FIS2015-67963-P), the Grant Agency of
the Czech Republic (Grant No. 15-03194S), and the IGA Project of the
Palack{\'y} University (Grant No. IGA PrF 2016-005).

\appendix

\section{Optimal estimators for homodyne tomography}
\label{app:optimality}

\subsection{First-moment estimation}
\label{subsec:app:firstmom-est}

In this discussion, the reconstruction accuracy of the estimator $\widehat{\rvec{r}}_\textsc{hom}$ for $\rvec{r}$ shall be taken to be the usual MSE distance measure
\begin{equation}
\mathcal{D}_\textsc{1,hom}=\overline{\left(\widehat{\rvec{r}}_\textsc{hom}-\rvec{r}\right)^2}
\end{equation}
that is typically defined for columns. One straightforward way to obtain an estimator $\widehat{\rvec{r}}_\textsc{hom}$ is to make use of $\left<X_\vartheta\right>=\left<X\right>\cos\vartheta+\left<P\right>\sin\vartheta$ to ascertain that
\begin{equation}
\dyadic{L}_\vartheta\rvec{r}=\rvec{r}_\vartheta
\end{equation}
for an $n_\vartheta\times2$ matrix $\dyadic{L}_\vartheta$ ($n_\vartheta$ being the number of bins for the LO phases $\vartheta$) and a column $\rvec{r}_\vartheta$ of $n_\vartheta$ true averages $\left<X_\vartheta\right>$. The highly overcomplete nature of the measurement thus permits us to define, for any experimentally obtained estimates of average values $\rvec{r}_\vartheta\equiv\left(\widehat{\left<X_1\right>}\,\,\widehat{\left<X_2\right>}\,\,\ldots\,\, \widehat{\left<X_{n_\vartheta}\right>}\right)^\textsc{t}$ $\left(\overline{\widehat{\left<X_k\right>}}=\left<X_{\vartheta_k}\right>\right)$,
\begin{equation}
\widehat{\rvec{r}}^{(\textsc{lin})}_\textsc{hom}=\dyadic{L}_\vartheta^{-}\left<\rvec{R}_\vartheta\right>
\label{eq:lin_est_firstmom}
\end{equation}
as the linear estimator of interest using the pseudoinverse $\dyadic{L}_\vartheta^{-}$ of $\dyadic{L}_\vartheta$. This estimator, however, is suboptimal in the sense that it does \emph{not} minimize the MSE $\mathcal{D}_{1,\textsc{hom}}$.

To obtain the best estimator for $\rvec{r}$ [often known as the linear unbiased estimator (BLUE)] that minimizes the MSE, we resort to the linear optimization of
\begin{equation}
\widehat{\rvec{r}}_\textsc{hom}=\sum^{n_\vartheta}_{k=1}\rvec{v}_k\widehat{\left<X_k\right>}
\label{eq:gen_est_firstmom}
\end{equation}
over all possible \emph{reconstruction columns} $\rvec{v}_k$ for the estimates $\widehat{\left<X_k\right>}$. Data consistency according to $\widehat{\left<X_k\right>}=\rvec{u}_k^{\textsc{t}}\widehat{\rvec{r}}_\textsc{hom}$
requires these reconstruction columns, or \emph{dual columns}, to satisfy the property
\begin{equation}
\sum^{n_\vartheta}_{k=1}\rvec{v}_k\rvec{u}_k^{\textsc{t}}=\dyadic{1}=\sum^{n_\vartheta}_{k=1}\rvec{u}_k\rvec{v}_k^{\textsc{t}}
\label{eq:dual_columns}
\end{equation}
with the measurement columns $\rvec{u}_k=\rvec{u}_{\vartheta_k}=\TP{(\cos\vartheta_k\,\,\sin\vartheta_k)}$. Logically, we must have
\begin{equation}
\overline{\widehat{\rvec{r}}_\textsc{hom}}=\sum^{n_\vartheta}_{k=1}\rvec{v}_k\left<X_k\right>=\rvec{r}\,.
\label{eq:gen_true_firstmom}
\end{equation}

The Lagrange function for the optimization is therefore
\begin{align}
\mathcal{L}_{\textsc{hom}}=\mathcal{D}_\textsc{hom}-\Tr{\dyadic{\Lambda}\left(\sum^{n_\vartheta}_{k=1}\rvec{u}_k\rvec{v}_k^{\textsc{t}}-\dyadic{1}\right)}\,,
\end{align}
where $\dyadic{\Lambda}$ is the Lagrange matrix for the dual-column constraint in \eqref{eq:dual_columns}. In terms of the dual columns,
\begin{align}
\mathcal{D}_\textsc{1,hom}=&\,\sum^{n_\vartheta}_{k=1}\sum^{n_\vartheta}_{k'=1}\rvec{v}_k^\textsc{t}\rvec{v}_{k'}\left(\overline{\widehat{\left<X_k\right>}\widehat{\left<X_{k'}\right>}}-\left<X_k\right>\left<X_{k'}\right>\right)\nonumber\\
=&\,\sum^{n_\vartheta}_{k=1}\rvec{v}_k^\textsc{t}\rvec{v}_{k}\overline{\widehat{\left<X_k\right>}^2}+\sum_{k\neq k'}\rvec{v}_k^\textsc{t}\rvec{v}_{k'}\overline{\widehat{\left<X_k\right>}}\,\,\overline{\widehat{\left<X_{k'}\right>}}\nonumber\\
&\,-\sum^{n_\vartheta}_{k=1}\sum^{n_\vartheta}_{k'=1}\rvec{v}_k^\textsc{t}\rvec{v}_{k'}\left<X_k\right>\left<X_{\vartheta_{k'}}\right>\,.
\end{align}
Since the unbiased estimate
\begin{equation}
\widehat{\left<X_k\right>}=\dfrac{1}{N_k}\sum^{n_x}_{j=1}n_{jk}x_{jk}
\end{equation}
is an average sum of all the measured $n_x$ voltage readings $x_{jk}$ per LO phase that are distributed according to the multinomial distribution of random multinomial weights $\sum_jn_{jk}=N_k$, the second moment is given by
\begin{align}
\overline{\widehat{\left<X_k\right>}^2}&=\dfrac{1}{N_k^2}\sum^{n_x}_{j=1}\sum^{n_x}_{j'=1}\overline{n_{jk}n_{j'k}}x_{jk}x_{j'k}\nonumber\\
&=\dfrac{1}{N_k}\sum^{n_x}_{j=1}p_{jk}x_{jk}^2+\dfrac{N_k-1}{N_k}\sum^{n_x}_{j=1}\sum^{n_x}_{j'=1}p_{jk}p_{j'k}x_{jk}x_{j'k}\nonumber\\
&=\dfrac{1}{N_k}\left<X_k^2\right>+\dfrac{N_k-1}{N_k}\left<X_k\right>^2\,,
\end{align}
The final equality is valid for sufficiently large data (bins) for all phases, as $p_{jk}\rightarrow\D x_\vartheta\, p(x_\vartheta,\vartheta)$ and
\begin{align}
\sum^{N_k}_{j=1}p_{jk}x_{jk}^2&\rightarrow\int\D x_\vartheta \,p(x_\vartheta,\vartheta)\,x_\vartheta^2\nonumber\\
&=\int\D x_\vartheta \left<\ket{x_\vartheta}\bra{x_\vartheta}\right>x_\vartheta^2=\left<X^2_\vartheta\right>\,.
\end{align}
So, we finally get
\begin{equation}
\mathcal{D}_{1,\textsc{hom}}=\sum^{n_\vartheta}_{k=1}\dfrac{\rvec{v}_k^\textsc{t}\rvec{v}_{k}}{N_k}\left(\left<X_k^2\right>-\left<X_k\right>^2\right)\,.
\label{eq:D_hom_firstmom}
\end{equation}
A simple variation of $\mathcal{L}_\textsc{hom}$ therefore gives
\begin{align}
\updelta\mathcal{L}_\textsc{hom}=&\,\sum^{n_\vartheta}_{k=1}\dfrac{\updelta\rvec{v}_k^\textsc{t}\rvec{v}_{k}+\rvec{v}_k^\textsc{t}\updelta\rvec{v}_{k}}{N_k}\left(\left<X_k^2\right>-\left<X_k\right>^2\right)\nonumber\\
&-\dfrac{1}{2}\Tr{\dyadic{\Lambda}\sum^{n_\vartheta}_{k=1}\left(\rvec{u}_k\updelta\rvec{v}_k^{\textsc{t}}+\updelta\rvec{v}_k\rvec{u}_k^{\textsc{t}}\right)}\equiv0\,,
\end{align}
or
\begin{align}
\dfrac{1}{2}\dyadic{\Lambda}&=\dyadic{F}\!\left(\{\left<X_k\right>,\left<X_k^2\right>\}\right)\equiv\sum^{n_\vartheta}_{k=1}\rvec{u}_{k}\rvec{u}_{k}^\textsc{t}\dfrac{N_k}{\left<X_k^2\right>-\left<X_k\right>^2}\,,\nonumber\\
\rvec{v}_k&=\dfrac{N_k}{\left<X_k^2\right>-\left<X_k\right>^2}\dyadic{F}\!\left(\{\left<X_k\right>,\left<X_k^2\right>\}\right)^{-1}\rvec{u}_k
\end{align}
The matrix $\dyadic{F}\!\left(\{\left<X_k\right>,\left<X_k^2\right>\}\right)$ is known as the frame matrix.

The BLUE therefore depends on the true moments which are certainly unavailable in the first place, for no tomography is otherwise necessary at all. Nonetheless, one can substitute the estimated moments for them to obtain an asymptotically efficient optimal estimator that approximates the BLUE. An unbiased estimate for the second moment is given by
\begin{equation}
\widehat{\left<X^2_k\right>}=\dfrac{1}{N_k}\sum^{N_k}_{j=1}n_{jk}x^2_{jk}\,,
\label{eq:secmom_est}
\end{equation}
so that the asymptotically optimal estimator is given by
\begin{align}
\widehat{\rvec{r}}^{(\textsc{opt})}_\textsc{hom}&=\dyadic{W}_1^{-1}\sum^{n_\vartheta}_{k=1}\rvec{u}_k\dfrac{N_k\widehat{\left<X_k\right>}}{\widehat{\left<X^2_k\right>}-\widehat{\left<X_k\right>}^2}\,\nonumber\\
\dyadic{W}_1&=\sum^{n_\vartheta}_{k=1}\rvec{m}_{k}\dfrac{N_k}{\widehat{\left<X_k^2\right>}-\widehat{\left<X_k\right>}^2}\,,
\end{align}

It is easy to see that when the estimated moments approach the true moments, this optimal estimator attains the sCRB. Directly from Eq.~\eqref{eq:D_hom_firstmom}, we immediately know that its corresponding MSE is given by
\begin{equation}
\mathcal{D}^{(\textsc{opt})}_{1,\textsc{hom}}=\Tr{\dyadic{F}\!\left(\{\left<X_k\right>,\left<X_k^2\right>\}\right)^{-1}}
\end{equation}
and all we need to realize is that for sufficiently large $N$ and uniformly-distributed quadrature outcomes, $N_k/N\rightarrow\D\vartheta/\pi$ and the frame matrix
\begin{equation}
\dfrac{1}{N}\dyadic{F}\!\left(\{\left<X_k\right>,\left<X_k^2\right>\}\right)\rightarrow \int_{(\pi)}\dfrac{\D\vartheta}{\pi}\dfrac{\dyadic{m}_\vartheta}{\left<X_{\vartheta}^2\right>-\left<X_{\vartheta}\right>^2}=\FHOMoneN
\end{equation}
is nothing more than the Fisher matrix introduced in Eq.~\eqref{eq:Fisher_HOM_first}. This also means that the BLUE and the asymptotically optimal estimator are both asymptotically as efficient as the ML estimator.

This construction comes with a basic and important lesson. The simple linear estimator $\widehat{\rvec{r}}^{(\textsc{lin})}_\textsc{hom}$ in Eq.~\eqref{eq:lin_est_firstmom}, which is suboptimal, depends only on the first moments. To improve the reconstruction accuracy, more aspects of the data that are attributed to the figure of merit chosen to measure this accuracy would have to be incorporated systematically. In the case of the MSE, these are linear combinations of both the first and second moments, or at least their estimates. Put differently, we should always use the reconstruction estimator that optimize the figure of merit we choose to rank the goodness of the reconstruction.

\begin{table*}[htp]
	\begin{ruledtabular}		
		\begin{small}
			\resizebox{2\columnwidth}{!}{%
				\begin{tabular}{lll}
					\vspace{1ex}
					Class of Quantum States & Quadrature Characteristic Function & Husimi Characteristic Function\\ \hline
					{\bf Gaussian} & $\displaystyle\exp\left(-\frac{1}{2}\left(\TP{\rvec{u}}_\vartheta\,\dyadic{G}\,\rvec{u}_\vartheta\right)^{\!2} k^2+\I\,\TP{\rvec{u}}_\vartheta\rvec{r}_0\,k\right)$ \vphantom{$\dfrac{{{W^W}^W}^W}{{{W^W}^W}^W}$} & $\displaystyle\E{g^*\alpha_0+g\alpha^*_0}\,\exp\!\left(\frac{\DET{\dyadic{G}_\textsc{het}}}{2}\rvec{g}^\dagger\dyadic{M}\,\rvec{g}\right)$ \vphantom{$\dfrac{{{W^W}^W}^W}{{{W^W}^W}^W}$}\\[3ex] 
					{\bf Fock} & $\displaystyle\E{-\frac{k^2}{4}}\LAG{n}{\frac{k^2}{2}}$ & $\displaystyle\pFq{1}{1}{n+1}{1}{|g|^2}$\\[3ex]
					{\bf Even/odd coherent} & $\displaystyle\E{-\frac{k^2}{4}}\dfrac{\cos(k x_\vartheta)\pm\E{-2|\alpha_0|^2}\cosh(k p_\vartheta)}{1\pm\E{-2|\alpha_0|^2}}$ & $\displaystyle
					\dfrac{\E{-|\alpha_0|^2}}{2\pm2\,\E{-2|\alpha_0|^2}}\left[\begin{matrix*}[l]
					&\E{|g+\alpha_0|^2}+\E{|g-\alpha_0|^2}\\
					\pm\!\!\!\!&\E{(g^*-\alpha_0^*)(g+\alpha_0)}\pm\,\text{c.c.}
					\end{matrix*}\right]
					$\\[3ex]
					{\bf Displaced Fock} & $\displaystyle\E{-\frac{k^2}{4}+\I kx_\vartheta}\LAG{m}{\frac{k^2}{2}}$ & $\displaystyle\E{g^*\alpha_0+g\alpha^*_0}\pFq{1}{1}{m+1}{1}{|g|^2}$\\[3ex]
					{\bf Photon-added coherent} & $\displaystyle\E{\frac{k^2}{4}}\dfrac{\pFq{1}{1}{m+1}{1}{\left(\alpha_0+\frac{\I k}{\sqrt{2}}\E{\I\vartheta}\right)\left(\alpha_0^*+\frac{\I k}{\sqrt{2}}\E{-\I\vartheta}\right)}}{\pFq{1}{1}{m+1}{1}{|\alpha_0|^2}}$ & $\displaystyle\dfrac{\pFq{1}{1}{m+1}{1}{|g+\alpha_0|^2}}{\pFq{1}{1}{m+1}{1}{|\alpha_0|^2}}$
				\end{tabular}
			}
		\end{small}
	\end{ruledtabular}	
	\caption{\label{tab:char} A list of characteristic functions for all the quantum states discussed. The symbols in this table are defined as $\alpha_0\E{-\I\vartheta}=(x_\vartheta+\I p_\vartheta)/\sqrt{2}$ where $x_0=x_{\vartheta=0}$ and $p_0=p_{\vartheta=0}$, $\rvec{g}=\TP{(-g\,\,\,g^*)}$, $\dyadic{M}=\dyadic{H}^\dagger\dyadic{G}_\textsc{het}^{-1}\,\dyadic{H}$, and $\dyadic{H}\,\widehat{=}\dfrac{1}{\sqrt{2}}\begin{pmatrix}
		1 & 1\\
		-\I & \I
		\end{pmatrix}$.}
\end{table*}

\subsection{Second-moment estimation}
\label{subsec:app:secmom-est}

By the same token, we can construct the optimal estimator that approximates the BLUE for second-moment estimation by minimizing the MSE
\begin{equation}
\mathcal{D}_{2,\textsc{hom}}=\overline{\Tr{\left(\widehat{\dyadic{G}}_{2,\textsc{hom}}-\dyadic{G}_{2,\textsc{hom}}\right)^2}}
\end{equation}
over the estimator that is of the linear form
\begin{equation}
\widehat{\dyadic{G}}_{2,\textsc{hom}}=\sum^{n_\vartheta}_{k=1}\dyadic{\Theta}_k\widehat{\left<X_{k}^2\right>}
\end{equation}
with respect to the second-moment estimates. This form is a natural extension to the column estimator $\widehat{\rvec{r}}_\textsc{hom}$ \emph{via} a generalization of the dual columns $\rvec{v}_k$ to \emph{dual} matrices $\dyadic{\Theta}_k$. Completely analogous to the discussion in Appendix~\ref{subsec:app:firstmom-est}, consistency with $\widehat{\left<X^2_k\right>}=\rvec{u}_k^{\textsc{t}}\widehat{\dyadic{G}}_{2,\textsc{hom}}\rvec{u}_k$ implies that
\begin{equation}
\widehat{\dyadic{G}}_{2,\textsc{hom}}=\sum^{n_\vartheta}_{k=1}\dyadic{\Theta}_k\rvec{u}_k^{\textsc{t}}\widehat{\dyadic{G}}_{2,\textsc{hom}}\rvec{u}_k\,.
\label{eq:G2hom_consistency}
\end{equation}
The above relation can be simplified by introducing the \emph{vectorization} notation $\VEC{\dyadic{Y}}$ that turns a matrix $\dyadic{Y}$ into a column. Since all two-dimensional matrices considered here are real and symmetric, they are essentially characterized by three real parameters. Hence in our context, given that
\begin{equation}
\dyadic{Y}\,\widehat{=}\begin{pmatrix}
y_1 & y_2\\
y_2 & y_3
\end{pmatrix}\,,
\end{equation}
the vectorized quantity is defined as
\begin{equation}
\VEC{\dyadic{Y}}\,\widehat{\equiv}\begin{pmatrix}
y_1\\
\sqrt{2}\,y_2\\
y_3
\end{pmatrix}\,.
\end{equation}
This operation is a variant of the usual column-stacking vectorization operation to apply on $2\times2$ real symmetric matrices for our case to make contact with the property $\Tr{\dyadic{Y}_1\dyadic{Y}_2}=\TP{\VEC{\dyadic{Y}_1}}\VEC{\dyadic{Y}_2}$ between any pair of such matrices $\dyadic{Y}_1$ and $\dyadic{Y}_2$. In this notation, Eq.~\eqref{eq:G2hom_consistency} becomes
\begin{equation}
\VEC{\widehat{\dyadic{G}}_{2,\textsc{hom}}}=\sum^{n_\vartheta}_{k=1}\VEC{\dyadic{\Theta}_k}\VEC{\dyadic{m}_k}^\textsc{t}\VEC{\widehat{\dyadic{G}}_{2,\textsc{hom}}}\,,
\end{equation}
which is equivalent to the vectorized constraint
\begin{equation}
\sum^{n_\vartheta}_{k=1}\VEC{\dyadic{\Theta}_k}\VEC{\dyadic{m}_k}^\textsc{t}=\dyadic{1}=\sum^{n_\vartheta}_{k=1}\VEC{\dyadic{m}_k}\VEC{\dyadic{\Theta}_k}^\textsc{t}\,.
\end{equation}

As usual, to derive the expression for the optimal estimator, we first calculate $\mathcal{D}_{2,\textsc{hom}}$ in terms of the dual matrices. For this we shall need the average of the square of the estimate $\widehat{\left<X_{k}^2\right>}$ defined in Eq.~\eqref{eq:secmom_est}:
\begin{equation}
\overline{\widehat{\left<X_{k}^2\right>}^2}=\dfrac{1}{N_k}\left<X_k^4\right>+\dfrac{N_k-1}{N_k}\left<X_k^2\right>^2\,,
\end{equation}
from which gives the expression
\begin{equation}
\mathcal{D}_{2,\textsc{hom}}=\sum^{n_\vartheta}_{k=1}\dfrac{\VEC{\dyadic{\Theta}}_k^\textsc{t}\VEC{\dyadic{\Theta}}_{k}}{N_k}\left(\left<X_k^4\right>-\left<X_k^2\right>^2\right)
\label{eq:D_hom_secmom}
\end{equation}
for the MSE. Then, by carrying out the variation of the appropriate Lagrange function similar to the calculations in Appendix~\eqref{subsec:app:firstmom-est} and remembering the additional association $\dyadic{M}_k=\VEC{\dyadic{m}_{k}}\VEC{\dyadic{m}_{k}}^\textsc{t}$, we find that the optimal matrices for the BLUE are
\begin{align}
\dyadic{F}\!\left(\{\left<X_k^2\right>,\left<X_k^4\right>\}\right)\equiv\sum^{n_\vartheta}_{k=1}\dyadic{M}_k\dfrac{N_k}{\left<X_k^4\right>-\left<X_k^2\right>^2}\,,\nonumber\\
\VEC{\dyadic{\Theta}_k}=\dfrac{N_k}{\left<X_k^4\right>-\left<X_k^2\right>^2}\dyadic{F}\!\left(\{\left<X_k^2\right>,\left<X_k^4\right>\}\right)^{-1}\VEC{\dyadic{m}_k}\,.
\end{align}
Finally, the asymptotically optimal estimator is given by
\begin{align}
\widehat{\dyadic{G}}^{(\textsc{opt})}_{2,\textsc{hom}}&=\dyadic{W}_2^{-1}\sum^{n_\vartheta}_{k=1}\VEC{\dyadic{m}_k}\dfrac{N_k\widehat{\left<X_k^2\right>}}{\widehat{\left<X^4_k\right>}-\widehat{\left<X_k^2\right>}^2}\,,\nonumber\\
\dyadic{W}_2&=\sum^{n_\vartheta}_{k=1}\dyadic{M}_k\dfrac{N_k}{\widehat{\left<X_{k}^4\right>}-\widehat{\left<X_{k}^2\right>}^2}\,.
\end{align}
That this estimator asymptotically attains the sCRB for second-moment estimation is again clear.

\section{List of characteristic functions}
\label{app:char_func}

In calculating the moments for both the homodyne and heterodyne schemes, it is extremely useful to start with the relevant characteristic functions for both schemes. To facilitate the discussions in the main article, we have supplied a list of quadrature characteristic functions $\left(\left<\E{\I k X_\vartheta}\right>\right)$ for the homodyne scheme and a list of Husimi characteristic functions $\displaystyle\left[\overline{\E{g^*\alpha+g\alpha^*}}\,,\,\,g=(u+\I v)/\sqrt{2}\right]$ for the heterodyne scheme respectively in Table~\ref{tab:char} in this appendix section. Then the two kinds of moments can then be readily computed by the prescriptions
\begin{align}
\left<X^m_\vartheta\right>&=\left(-\I\dfrac{\partial}{\partial k}\right)^m\left<\E{\I k X_\vartheta}\right>\Bigg|_{k=0}\,,\nonumber\\
\overline{x^kp^l}&=\left(\dfrac{\partial}{\partial u}\right)^k\left(\dfrac{\partial}{\partial v}\right)^l\overline{\E{g^*\alpha+g\alpha^*}}\Bigg|_{u,v=0}\,,
\end{align} 
which simply involves multiple differentiations with respect to the free variables and later setting these variables to zero. Some useful identities for the confluent hypergeometric functions and Laguerre polynomials that allow for consistency verification between two characteristic functions of different quantum states are given below:
\begin{align}
\LAG{0}{x}&=1\,,\nonumber\\
\LAG{1}{x}&=1-x\,,\nonumber\\
\pFq{1}{1}{1}{1}{x}&=\E{x}\,,\nonumber\\
\pFq{1}{1}{2}{1}{x}&=\E{x}(1+x)\,,\nonumber\\
\pFq{1}{1}{n+1}{1}{-x}&=\E{-x}\LAG{n}{x}\,.
\end{align}

\end{document}